\documentclass[10pt,conference,letterpaper]{IEEEtran}
\usepackage{times,amsmath,epsfig,amssymb,latexsym,multirow}

\usepackage[utf8]{inputenc}

\usepackage{boxedminipage}

\usepackage{listings}

\usepackage{xspace}

\usepackage{caption}

\usepackage{cleveref}
\usepackage{algorithm, algorithmic}

\usepackage{subfigure}
\usepackage{graphicx}

\newtheorem{definition}{Definition}

\newtheorem{theorem}{Theorem}

\newtheorem{guideline}[theorem]{Guideline}

\newcommand{\todo}[1]{\textbf{Todo:}\{ \emph{#1} \}}
\newcommand{\discuss}[1]{\textbf{Discuss:}\{ \emph{#1} \}}
\newcommand{\comment}[1]{}

\renewcommand{\AA}{\ensuremath{\mathcal{A}}}

\newcommand{\MM}{\ensuremath{\mathcal{M}}}

\newcommand{\UG}{\ensuremath{\mathsf{UG}}\xspace}			% the flat grid method
\newcommand{\AG}{\ensuremath{\mathsf{AG}}\xspace}			% the adaptive-grids method
\newcommand{\DAG}{\ensuremath{\mathsf{DAG}}\xspace}			% the differentiated-adaptive-grids method

\newcommand{\Lap}[1]{\mathsf{Lap}\left(#1\right)\xspace}

\newcommand{\VMax}{\mathsf{V_{Max}}\xspace}
\newcommand{\VUnit}{\mathsf{V_{Unit}}\xspace}

\newcommand{\mypara}[1]{\vspace*{0.05in}\noindent\textbf{#1}$\;$}

\renewcommand{\Pr}[1]{\ensuremath{\mathsf{Pr}\left[#1\right]}\xspace}
\newcommand{\myexp}[1]{\ensuremath{e^{#1}}\xspace}

\newcommand{\DP}[1]{\ensuremath{#1\mbox{-}\mathsf{DP}}\xspace}
\newcommand{\eDP}{\DP{\epsilon}}

\title{Differentially Private Grids for Geospatial Data}
\author{%
{Wahbeh Qardaji, Weining Yang, Ninghui Li}%
\vspace{1.6mm}\\
\fontsize{10}{10}\selectfont\itshape
Department of Computer Science, Purdue University\\
305 N. University Street, West Lafayette, IN 47907, USA\\
\fontsize{9}{9}\selectfont\ttfamily\upshape
\{wqardaji, yang469, ninghui\}@cs.purdue.edu%
}
\begin{document}
\maketitle
\begin{abstract}
In this paper, we tackle the problem of constructing a differentially private synopsis for two-dimensional datasets such as geospatial datasets.  
The current state-of-the-art methods work by performing recursive binary partitioning of the data domains, and constructing a hierarchy of partitions.
We show that the key challenge in partition-based synopsis methods lies in choosing the right partition granularity to balance the noise error and the non-uniformity error.  
We study the uniform-grid approach, which applies an equi-width grid of a certain size over the data domain and then issues independent count queries on the grid cells.  This method has received no attention in the literature, probably due to the fact that no good method for choosing a grid size was known.  Based on an analysis of the two kinds of errors, we propose a method for choosing the grid size.  Experimental results validate our method, and show that this approach performs as well as, and often times better than, the state-of-the-art methods.
%We also analyze why hierarchical methods provide limited benefit in 2-dimensional data, and analytically show that for 3 or higher dimensional datasets, hierarchical methods likely perform even worse when compared with the uniform-grid method.

We further introduce a novel adaptive-grid method.  The adaptive grid method lays a coarse-grained grid over the dataset, and then further partitions each cell according to its noisy count. Both levels of partitions are then used in answering queries over the dataset.  
This method exploits the need to have finer granularity partitioning over dense regions and, at the same time, coarse partitioning over sparse regions.
Through extensive experiments on real-world datasets, we show that this approach consistently and significantly outperforms the uniform-grid method and other state-of-the-art methods.
\end{abstract}

% NOTE keywords are not used for conference papers so do not populate them
\section{Introduction} \label{sec:intro}
%!TEX root = ./md_hist.tex

We interact with location-aware devices on a daily basis. Such devices range from GPS-enabled cell-phones and tablets, to navigation systems. Each device can report a multitude of location data to centralized servers. Such location information, commonly referred to as geospatial data, can have tremendous benefits if properly processed and analyzed. For many businesses, a location-based view of information can enhance business intelligence and enable smarter decision making. For many researchers, geospatial data can add an interesting dimension. Location information from cell-phones, for instance, can help in various social research that is interested in how populations settle and congregate. Furthermore, location from in-car navigation systems can help provide information on areas of common traffic congestion.

If shared, such geo-spatial data can have significant impact for research and other uses. Sharing such information, however, can have significant privacy implications. In this paper, we study the problem of releasing static geo-spatial data in a private manner. In particular, we introduce methods of releasing a synopsis of two-dimensional datasets while satisfying differential privacy.

Differential privacy~\cite{Dwo06} has recently become the defacto standard for privacy preserving data release, as it is capable of providing strong worst-case privacy guarantees. We consider two-dimensional, differentially private, synopsis methods in the following framework. Given a dataset and the two-dimensional domain that tuples in the dataset are in, we view each tuple as a point in two-dimensional space. One partitions the domain into cells, and then obtains noisy counts for each cell in a way that satisfies differential privacy. The differentially private synopsis consists of the boundaries of these cells and their noisy counts. This synopsis can then be used either for generating a synthetic dataset, or for answering queries directly.

In general, when answering queries, there are two sources of error in such differentially private synopsis methods. The first source is the noise added to satisfy differential privacy. This noise has a predefined variance and is independent of the dataset, but depends on how many cells are used to answer a query. The second source is the nature of the dataset itself. When we issue a query which only partially intersects with some cell, then we would have to estimate how many data points are in the intersected cells, assuming that the data points are distributed uniformly. The magnitude of this error depends both on the distribution of points in the dataset and on the partitioning. Our approach stems from careful examination of how these two sources of error depend on the grid size.

%Before introducing our approach, we look at existing methods used to apply differential privacy to this domain. 
Several recent papers have attempted to develop such differentially private synopsis methods for two-dimensional datasets~\cite{XXY10, CPS+12}. These papers adapt spatial indexing methods such as quadtrees and kd-trees to provide a private description of the data distribution.
 %We argue that there are various limitations to the efficacy of such methods. 
These approaches can all be viewed as adapting the binary hierarchical method, which works well for 1-dimensional datasets, to the case of 2 dimensions. The emphasis is on how to perform the partitioning, and the result is a deep tree. 

Somewhat surprisingly, none of the existing papers on summarizing multi-dimensional datasets compare with the simple uniform-grid method, which applies an equi-width $m\times m$ grid over the data domain and then issues independent count queries on the grid cells.  We believe one reason is that the accuracy of \UG is highly dependent on the grid size $m$, and how to choose the best grid size was not known.  We propose choosing $m$ to be $\sqrt{\frac{N\epsilon}{c}},$ where $N$ is the number of data points, $\epsilon$ is the total privacy budget, and $c$ is some small constant depending on the dataset.  Extensive experimental results, using $4$ real-world datasets of different sizes and features, validate our method of choosing $m$.
Experimental results also suggest that setting $c=10$ work well for datasets of different sizes and different choices of $\epsilon$, and show that \UG performs as well as, and often times better than the state-of-the-art hierarchical methods in~\cite{XXY10, CPS+12}.

This result is somewhat surprising, as hierarchical methods have been shown to greatly outperform the equivalence of uniform-grid in $1$-dimensional case~\cite{HRMS10,XWG11}.  We thus analyze the effect of dimensionality on the effectiveness of using hierarchies.

%Because of this, we adopt the approach of starting from the Uniform Grid method and trying to improve upon this method.  The side effect is that the emphasis in designing the algorithm shifts from choosing the location for partition, to choosing the partition granularity. Hence, we take a deeper look into the uniform grid approach, and we show how to choose the grid size to obtain better accuracy. Using experimental evaluation, we show that this simple method provides smaller error when answering queries.

We further introduce a novel adaptive-grid method.  This method is motivated by the need to have finer granularity partitioning over dense regions and, at the same time, coarse partitioning over sparse regions. The adaptive grid method lays a coarse-grained grid over the dataset, and then further partitions each cell according to its noisy count. Both levels of partitions are then used in answering queries over the dataset. We propose methods to choose the parameters for the partitioning by careful analysis of the aforementioned sources of error. %The differentiated adaptive grid method groups adjacent sparse cells and issues a coarser-grained count query over their union, thus most effectively utilizing the privacy budget. For both methods we analyze how to best chose the parameters for the partitioning.
%
%Through experiments, we validate our method for choosing the grid size for the uniform-grid method and compare the spatial indexing methods to the uniform grids approach.  We show that by carefully choosing parameters for the uniform grids approach, this simple and efficient method can perform as well as and often better than the spatial indexing techniques.
 %We then analyze the effect of hierarchical transformation approaches, including hierarchical constrained inference and wavelet transforms, on the uniform grids approach. We show that, while there are some benefits to using such approaches, such benefits are marginal.
Extensive experiments validate our methods for choosing the parameters, and show that the adaptive-grid method consistently and significantly outperforms the uniform grid method and other state-of-the-art methods.

The contributions of this paper are as follows:
\begin{enumerate}
 \item 
We identify that the key challenge in differentially private synopsis of geospatial datasets is how to choose the partition granularity to balance errors due to two sources, and propose a method for choosing grid size for the uniform grid method, based on an analysis of how the errors depend on the grid size.

 \item
We propose a novel, simple, and effective adaptive grid method, together with methods for choosing the key parameters. 

 \item
We conducted extensive evaluations using 4 datasets of different sizes, including geo-spatial datasets that have not been used in differentially private data publishing literature before.  Experimental results validate our methods and show that they outperform existing approaches.

 \item
We analyze why hierarchical methods do not perform well in $2$-dimensional case, and predict that they would perform even worse with higher dimensions.
\end{enumerate}

The rest of this paper is organized as follows. In Section~\ref{sec:probdef}, we set the scope of the paper by formally defining the problem of publishing two dimensional datasets using differential privacy. In Section~\ref{sec:related}, we discuss previous approaches and related work. We present our approach in Section~\ref{sec:approach}, and present the experimental results supporting our claims in Section~\ref{sec:results}. Finally, we conclude in Section~\ref{sec:conclusions}.

\comment{
A histogram is an important tool for summarizing data.  In recent years, significant progress has been made in the problem of publishing histograms over one-dimensional dataset in a differentially private way.  The utility goal is to ensure that range queries over the private histogram can be answered as accurately as possible.
The naive approach, which we call the flat method, is to issue a count query for each unit interval in the histogram. This is answered using the Laplacian mechanism introduced by Dwork et al.~\cite{Dwo06}.  Because each such query has sensitivity $1$, the magnitude of the added noise is small.  While this works well with histograms that have a small number of unit intervals, the error due to this method increases substantially as the histogram gets larger. In particular, when answering range queries using the private histogram, the error variance increases linearly in the size of the query range, resulting in a worst-case error standard deviation of $O(\sqrt{N})$ for histograms of size $N$.

Hay et al.~\cite{HRMS10} introduced the hierarchical method for optimizing differentially private histograms.  In this approach, in addition to asking for counts of unit-length intervals, one also asks for interval counts of larger granularity.  Conceptually, one can arrange all queried intervals into a tree, where the unit-length intervals are the leaves.  The benefit of this method is that a range query which includes many unit intervals can be answered using a small number of sub-intervals which exactly cover the query range.  The tradeoff over the flat method is that when queries are issued at multiple levels, each query must satisfy differential privacy for a smaller $\epsilon$.  We refer to this as the privacy budget allocated to (or consumed by) the query.  Hay et al.~\cite{HRMS10} also introduced constrained inference to improve over the basic hierarchical method, which exploits the observation that query results at different levels should satisfy certain consistency relationships to obtain improved estimates.

Xiao et al.~\cite{XWG11} introduced the Privlet method, which decomposes the original histogram using the Haar wavelet, then adds noise to the decomposed coefficients, and finally reconstructs the histogram using the noisy coefficients.  The benefit of this method is that, when answering range queries, noise added to the coefficients can be partially cancelled because of the nature of Haar wavelet processing.

When extending these methods to 2-dimensional or high-dimensional data, they may not work as well.
In this paper, we tackle the issue of publishing single and double dimension continuous data.

\cite{sp12} uses differentially private KD-Trees to publish spatial data. Building a differentially private KD-Tree involves recursive partitioning along the median of alternating dimensions until each region in the data domain contains a small number of points. While this approach works well in some situations, there are various limitations to its efficacy. \todo{complete}

Our approach builds on a simple observation: when data is uniformly distributed in the domain, one can answer queries accurately without the need to partition. This is because one would have accurate a priori knowledge of the number of points in each region of the domain. If, on the other hand, the distribution lacks uniformity, then one would need to partition in order to gain a better understanding of the underlying distribution. One would ideally stop partitioning when each region has a uniform distribution of points. We observe the regions with small counts tend to be uniformly distributed (for instance, a region with 1 or zero points is always uniformly distributed), hence it would be reasonable to stop partitioning when counts in each region are low. When using this to achieve differential privacy, one needs to consider how to partition while keeping the privacy budget in mind.

There a various possible methods to partition. Using the median is one such method. This, however, needs to be applied to each dimension separately. Furthermore, it cannot reliably produce regions of uniform density \discuss{it can eliminate regions with no points after a couple of levels}. Another reasonable method to capture our observation is as follows: first, coarsely divide the domain into sub-regions and then query each region. If the region is dense, partition further.
%The reason this works reasonably well in \cite{} is because it separates regions
}

\section{Problem Definition} \label{sec:probdef}
%!TEX root = ./md_hist.tex

%In this paper, we consider the case of differentially private analysis for multidimensional datasets.

\subsection{Differential Privacy} \label{sec:preliminaries:differential}

Informally, differential privacy requires that the output of a data analysis mechanism be approximately the same, even if any single tuple in the input database is arbitrarily added or removed.
\begin{definition}[{$\epsilon$-Differential Privacy~\cite{Dwo06,DMN+06}}] \label{def:diff}
A randomized mechanism $\AA$ gives $\epsilon$-differential privacy if for any pair of neighboring datasets $D$ and $D'$, and any $S\in \mathit{Range}(\AA)$,
%\begin{equation}\label{eqn:dp}
$$\Pr{\AA(D)=S} \leq e^{\epsilon}\cdot \Pr{\AA(D')=S}.$$
%\end{equation}
\end{definition}

In this paper we consider two datasets $D$ and $D'$ to be neighbors if and only if either $D=D' + t$ or $D'=D + t$, where $D +t$ denotes the dataset resulted from adding the tuple $t$ to the dataset $D$. We use $D\simeq D'$ to denote this.  This protects the privacy of any single tuple, because adding or removing any single tuple results in $e^{\epsilon}$-multiplicative-bounded changes in the probability distribution of the output.  If any adversary can make certain inference about a tuple based on the output, then the same inference is also likely to occur even if the tuple does not appear in the dataset.

%An alternative is to define $D$ and $D'$ to be neighbors when they contain the same number of tuples and all except for one are the same.  When the number of tuples in the input dataset is not sensitive, either of these two notions offers sufficient privacy protection.

Differential privacy is composable in the sense that combining multiple mechanisms that satisfy differential privacy for $\epsilon_1, \cdots,\epsilon_m$ results in a mechanism that satisfies $\epsilon$-differential privacy for $\epsilon=\sum_{i} \epsilon_i$.  Because of this, we refer to $\epsilon$ as the privacy budget of a privacy-preserving data analysis task.  When a task involves multiple steps, each step uses a portion of $\epsilon$ so that the sum of these portions is no more than $\epsilon$.

%There are several approaches for designing  mechanisms that satisfy $\epsilon$-differential privacy.  In this paper we use two of them.
To compute a function $g$ on the dataset $D$ in a differentially privately way, one can add to $g(D)$ a random noise drawn from the Laplace distribution.  The magnitude of the noise depends on $\mathsf{GS}_g$, the \emph{global sensitivity} or the $L_1$ sensitivity of $g$.  Such a mechanism $\AA_g$ is given below:
$$
\begin{array}{cl}
& \AA_g(D) =g(D)+\Lap{\frac{\mathsf{GS}_g}{\epsilon}}
  \\
\mbox{where} &  \mathsf{GS}_g  = \max_{(D,D') : D \simeq D'} |g(D) - g(D')|,
\\
\mbox{and}& \Pr{\Lap{\beta}=x}  = \frac{1}{2\beta} \myexp{-|x|/\beta}
\end{array}
$$
In the above, $\Lap{\beta}$ denotes a random variable sampled from the Laplace distribution with scale parameter $\beta$.  The variance of $\Lap{\beta}$ is $2\beta^2$; hence the standard deviation of $\Lap{\frac{\mathsf{GS}_g}{\epsilon}}$ is $\sqrt{2}\frac{\mathsf{GS}_g}{\epsilon}$.
%This is generally referred to as the \emph{Laplacian mechanism} for satisfying differential privacy.

\comment{
Without loss of generality, we assume that the domain for each attribute is $[0,1)$, that is the dataset $D\in [0,1)^d$.  We focus on the cases when $d$ is small, e.g., 1, 2, and 3.

We want to differentially private publish information so that range queries can be accurately answered.

A range query $r$ is given by $d$ intervals $[a_1,b_1),\cdots,[a_d,b_d)$, one for each dimension.  This query asks for the returns a count of all the tuples occurring between $a$ and $b$.  We consider the behaviors of different histogram publishing methods over a dataset $D$.  In the descriptions below, we leave the dataset $D$ implicit.

For a query $r$, we use $A(r)$ to denote the correct answer to $r$.  For a method $\MM$ and a query $r$, we use $Q_\MM(r)$ to denote the answer to the query $r$ when using the histogram constructed by method $\MM$ to answer the query $r$, and $E_{\MM}(r)$ to denote the absolute error for query $r$ under $\MM$.  That is
$$E_{\MM}(r)=|Q_\MM(r)-A(r)|.$$

Because the method $\MM$ is often randomized, $E_{\MM}(r)$ is a random variable.  We use the variance of $E_{\MM}(r)$ as a measurement of the error.  In particular, we consider the worst-case bound on the error variance.  Let $r_m$ be the query that maximizes the variance of $E_{\MM}(r)$.  We use $\VMax[\MM]$ to denote the variance of $E_{\MM}(r_m)$.  We will analyze the $\VMax[\MM]$ for different \MM.  The worst-case analysis gives a theoretical upper bound on the query accuracy.

In addition to the absolute error $E_{\MM}$, we also consider the relative error, $RE_{\MM}$ defined as
$$RE_{\MM}(r) = \frac{E_{\MM}(r)}{\max\{A(r), \rho\}}$$
where $\rho$ is a small constant. Using $\rho > 0$ avoids dividing by 0 when $A(r) = 0$. When the range of a query, $r$, is large, $RE_{\MM}(r)$ is likely to be very small since $A(r)$ is likely to be large. Hence, when one wants to bound the worst-case $RE_{\MM}(r)$, one needs to consider small ranges. In this case, we care about the variance of unit-length intervals in the histogram. We use $\VUnit[\MM]$ to denote the maximum variance of unit-length intervals when using \MM.
}

\subsection{Problem Definition} \label{sec:probdef:def}

We consider the following problem.  Given a 2-dimensional geospatial dataset $D$, our aim is to publish a synopsis of the dataset to accurately answer count queries over the dataset.
We consider synopsis methods in the following framework. Given a dataset and the two-dimensional domain that tuples in the dataset are in, we view each tuple as a point in two-dimensional space.
One partitions the domain into cells, and then obtains noisy counts for each cell in a way that satisfies differential privacy.  The differentially private synopsis consists of the boundary of these cells and their noisy counts. This synopsis can then be used either for generating a synthetic dataset, or for answering queries directly.

We assume that each query specifies a rectangle in the domain, and asks for the number of data points that fall in the rectangle. Such a count query can be answered using the noisy counts for cells in the following fashion.  If a cell is completely included in the query rectangle, then the noisy count is included in the total.  If a cell is partially included, then one estimates the point count in the intersection between the cell and the query rectangle, assuming that the points within the cell is distributed uniformly. For instance, if only half of the area of the cell is included in the query, then one assumes that half of the points are covered by the query.

\mypara{Two Sources of Error.}  Under this method, there are two sources of errors when answering a query.  The \textbf{noise error} is due to the fact that the counts are noisy.  To satisfy differential privacy, one adds, to each cell, an independently generated noise, and these noises have the same standard deviation, which we use $\sigma$ to denote.  When summing up the noisy counts of $q$ cells to answer a query, the resulting noise error is the sum of the corresponding noises.  As these noises are independently generated zero-mean random variables, they cancel each other out to a certain degree.  In fact, because these noises are independently generated, the variance of their sum equals the sum of their variances.  Therefore, the sum has variance $q\sigma^2$, corresponding to a standard deviation of $\sqrt{q} \sigma$.  That is, the noise error of a query grows linearly in $\sqrt{q}$.  Therefore, the finer granularity one partitions the domain into, the more cells are included in a query, and the larger the noise error is.

The second source of error is caused by cells that intersect with the query rectangle, but are not contained in it.  For these cells, we need to estimate how many data points are in the intersected cells assuming that the data points are distributed uniformly.  This estimation will have errors when the data points are not distributed uniformly.  We call this the \textbf{non-uniformity error}.  The magnitude of the non-uniformity error in any intersected cell, in general, depends on the number of data points in that cell, and is bounded by it.  Therefore, the finer the partition granularity, the lower the non-uniformity error.

As argued above, reducing the noise error and non-uniformity error imposes conflicting demands on the partition granularity.  The main challenge of partition-based differentially private synopsis lies in how to meet this challenge and reconcile the conflicting needs of noise error and non-uniformity error.

\section{Previous Approaches and Related Work} \label{sec:related}
%!TEX root = ./md_hist.tex

% Problem Definition

Differential privacy was presented in a series of papers \cite{DN03,DN04,BDMN05,DMN+06,Dwo06} and methods of satisfying it for evaluating some function over the dataset are presented in~\cite{Dwo06, NRS07, MT07}.

%\todo{Here I would like to discuss the three existing recursive partitioning approaches in SDM10, our paper, and ICDE12.  These approaches do binary partition, and use median method for partitioning.  Explain how they work.  Some approaches are generic, the ICDE12 paper is specifically for geospatial data.}

\mypara{Recursive Partitioning.}
Most approaches that directly address two-dimensional and spatial datasets use recursive partitioning~\cite{QL12, CPS+12, XXY10}.  These approaches perform a recursive binary partitioning of the data domain. %In each recursion, some dimension is chosen and then partitioning is performed along the median of that dimension.

Xiao et al.~\cite{XXY10} proposed adapting the standard spatial indexing method, KD-trees, to provide differential privacy. Nodes in a KD-tree are recursively split along some dimension. In order to minimize the non-uniformity error, Xiao et al. use the heuristic to choose the split point such that the two sub-regions are as close to uniform as possible.  %In addition, they implement a two-step algorithm that generates the kd-tree partitions based on the histogram generated from a cell partitioning in order to minimize the added noise.  \todo{I didn't understand this sentence; either comment out or clarify.}

Cormode et al.~\cite{CPS+12} proposed a similar approach. Instead of using a uniformity heuristic, they split the nodes along the median of the partition dimension. The height of the tree is predetermined and the privacy budget is divided among the levels. Part of the privacy budget is used to choose the median, and part is used to obtain the noisy count.
%
%This is used to query the median as well as a noisy count. Since privacy budget is divided among the levels and it used to determine the median, the resulting noise due to this method is high.
%
%In general, we find that adapting uniformity measures using differential privacy to be quite rigid and impractical. Furthermore, privacy budget is better spent issuing count queries. The splitting axis alternates at each level.  However,
% T via lines passing through the median data value
%; however, they adapt recursive partitioning for spatial data in specific.  which adapt standard spatial indexing methods such as KD-trees and Hilbert R-trees to provide a private description of the data distribution.  Since privacy budget is divided among the levels and it used to determine the median, the resulting noise due to this method would be high.
%
\cite{CPS+12} also proposed combining quad-trees with noisy median-based partitioning. In the quadtree, nodes are recursively divided into four equal regions via horizontal and vertical lines through the midpoint of each range. Thus no privacy budget is needed to choose the partition point.
%
%As opposed to the previous approaches, the privacy budget is just used to obtain noisy counts at each level of partitioning rather than to decide how to partition. Since the partitioning is data-independent, dense-regions would have large non-uniformity errors, while sparse regions have high noise errors.
The method that gives the best performance in~\cite{CPS+12} is a hybrid approach, which they call ``KD-hybrid''. This method uses a quadtree for the first few levels of partitions, and then uses the KD-tree approach for the other levels. A number of other optimizations were also applied in KD-hybrid, including the constrained inference presented in~\cite{HRMS10}, and optimized allocation of privacy budget.  Their experiments indicate that ``KD-hybrid'' outperforms the KD-tree based approach and the approach in~\cite{XXY10}.  

Qardaji and Li~\cite{QL12} proposed a general recursive partitioning framework for multidimensional datasets. At each level of recursion, partitioning is performed along the dimension which results in the most balanced partitioning of the data points. %The partitioning scheme at each level is determined using the exponential mechanism for differential privacy. %, which  emplys a quality function that incorporates several dimensions.
%At each level of recursion, part of the privacy is used to determine how to partition, and the rest is used to obtain a noisy count. Partitioning stops when the partitioned region has a small number of points.
The balanced partitioning employed by this method has the effect of producing regions of similar size.
%This, in turn reduces the relative error due to noise, since the resulting regions are non-sparse.
When applied to two-dimensional datasets, this approach is very similar to building a KD-tree based on noisy median.

%In Section~\ref{sec:approach}, we argue recursive partitioning approaches are limited in their efficacy.  We discuss the reason for this

We experimentally compare with the state-of-the-art KD-hybrid method.  In Section~\ref{sec:approach:idea}, we analyze the effect of dimensionality and show that hierarchical methods provide limited benefit in the $2$-dimensional case.

%In essence, the division of privacy budget among a large number of levels is undesirable as it results in large noise. In~Section~\ref{sec:results}, we show that our methods outperform KD-Hybrid.

\mypara{Hierarchical Transformations.} The recursive partitioning methods above essentially build a hierarchy over a representation of the data points. Several approaches have been presented in the literature to improve count queries over such hierarchies.

In \cite{HRMS10}, Hay et al.~proposed the notion of constrained inference for hierarchical methods to improve accuracy for range queries. This work has been mostly developed in the context of one-dimensional datasets. Using this approach, one would arrange all queried intervals into a binary tree, where the unit-length intervals are the leaves. Count queries are then issued at all the nodes in the tree.  Constrained inference exploits the consistency requirement that the parent's count should equal the sum of all children's counts to improve accuracy.
%To improve the accuracy of these counts, constrained inference is performed: for each set of parent and child nodes, one has two different counts for some interval. These two counts are combined such that a more accurate count is obtained.
%In corollary{sec:approach}, we argue recursive partitioning approaches are limited in their efficacy. In essence, the division of privacy budget among a large number of levels is undesirable as it results in large noise. Instead, we show how methods that employ a minimal number levels are better than hierarchical partitioning approaches.

In \cite{XWG11}, Xiao et al. propose the Privlet method answering histogram queries, which uses wavelet transforms. Their approach applies a Harr wavelet transform to the frequency matrix of the dataset. A Harr wavelet essentially builds a binary tree over the dataset, where each node (or ``coefficient'') represents the difference between the average value of the nodes in its right subtree, and the average value of the nodes in its left subtree. The privacy budget is divided among the different levels, and the method then adds noise to each transformation coefficient proportional to its sensitivity. These coefficients are then used to regenerate an anonymized version of the dataset by applying the reverse wavelet transformation. The benefit of using wavelet transforms is that they introduce a desirable noise canceling effect when answering range queries.

For two dimensional datasets, this method uses standard decomposition when applying the wavelet transform. Viewing the dataset as a frequency matrix, the method first applies the Harr wavelet transform on each row. The result is a vector of detail coefficients for each row. Then, using the matrix of detail coefficients as input, the method applies the transformation on the columns. Noise is then added to each cell, proportional to the sensitivity of the coefficient in that cell. To reconstruct the noisy frequency matrix, the method applies the reverse transformation on each column and then each row.
%Since this noise added to each matrix cell is effectively smaller than the standard hierarchical approaches, this method is expected to perform slightly better.
%However, since the privacy budget still has to be divided, the method has the same downfalls as the previously discussed methods.
%In \cite{HRMS10}, Hay et al. propose the notion of constrained inference for hierarchical methods to improve query accuracy for range queries. This work has been mostly developed in the context of one-dimensional datasets. Using this approach, one would arrange all queried intervals into a binary tree, where the unit-length intervals are the leaves. Count queries are then issued at all the nodes in the tree. To improve the accuracy of these counts, constrained inference is performed: for each set of parent and child nodes, one has two different counts for some interval. These two counts are combined such that a more accurate count is obtained. In \Cref{sec:general}, we show how this concept can also be applied to two-dimensional datasets.

%For two dimensions, this method requires further division of privacy budget as the wavelet transform is applied to each dimension. Hence, it has the same downfalls as the previously discussed methods.

Both constrained inference and wavelet methods have been shown to be very effective at improving query accuracy in the 1-dimensional case.  Our experiments show that applying them to a uniform grid provides small improvements for the 2-dimensional datasets.  We note that these methods can only be applied when one has decided what are the leaf cells.  When combined with the uniform grid method, it requires a method to choose the right grid size, as the performance will be poor when a wrong grid size is used.  In Section~\ref{sec:results}, we experimentally compare with the wavelet method. %can be applied after the smallest cells have been determined, and try to improve upon the method of obtaining noisy counts for just these cells.
%The effect is most pronounced for one-dimensional datasets. We show that applying either constrained inference or the Wavelet methods provide a small gain over

\mypara{Other related work.} Blum et al.~\cite{BLR08} proposed an approach that employs non-recursive partitioning, but their results are mostly theoretical and lack general practical applicability to the domain we are considering.

\cite{DNR+09, DRV10, LM12} provide methods of differentially private release which assume that the queries are known before publication. The most recent of such works by Li and Miklau~\cite{LM12} proposes the matrix mechanism. Given a workload of count queries, the mechanism automatically selects a different set of strategy queries to answer privately. It then uses those answers to derive answers to the original workload. %This approach, however, achieves a weaker version of differential privacy.
Other techniques for analyzing general query workloads under differential privacy have been discussed in \cite{HT10, LHR+10}. These approaches also require the base cells to be fixed.  Furthermore, they require the existence of a known set of queries, which are represented as a matrix, and then compute how to combine base cells to answer the original queries.  It is unclear how to use this method when one aims at answering arbitrary range queries.

 %\cite{FS10} gives an approach for data mining for differential privacy that uses the exponential mechanism; however, their approach is not concerned with publishing an entire dataset. %Furthermore,

A number of approaches exist for differentially private interactive data analysis, e.g.,~\cite{McS09}, and methods of improving the accuracy of such release~\cite{RR10} have been suggested. In such works, however, one interacts with a privacy aware database interface rather than getting access to a synopsis of the dataset. Our approach deals with the latter. %This classifies queries as ``easy'' or ``hard'', according to whether or not the majority of databases consistent with previous answers to hard queries would give an accurate answer to it. %(in which case the user already ``knows the answer'').
%A ``hard'' query is answered using the Laplacian mechanism. An ``easy'' query simply returns the corresponding median value. Our approach, however, deals with the non-interactive case.

\section{The Adaptive Partitioning Approach} \label{sec:approach}
%!TEX root = ./md_hist.tex

In this section, we present our proposed methods.

\subsection{The Uniform Grid Method - \UG} \label{sec:approach:ug}

Perhaps the simplest method one can think of is the Uniform Grid (\UG) method. This approach partitions the data domain into $m \times m$ grid cells of equal size, and then obtains a noisy count for each cell.  Somewhat surprisingly, none of the existing papers on summarizing multi-dimensional datasets compare with \UG.  We believe one reason is that the accuracy of \UG is highly dependent on the grid size $m$, and how to choose the best grid size was not known.

We propose the following guideline for choosing $m$ in order to minimize the sum of the two kinds of errors presented in Section~\ref{sec:probdef}.

\begin{guideline} \label{gl:size}
In order to minimize the errors due to \eDP \UG, the grid size should be about
$$\sqrt{\frac{N\epsilon}{c}},$$
where $N$ is the number of data points, $\epsilon$ is the total privacy budget, and $c$ is some small constant depending on the dataset.  Our experimental results suggest that setting $c=10$ works well for the datasets we have experimented with.
\end{guideline}

Below we present our analysis supporting this guideline.  As the sensitivity of the count query is $1$, the noise added for each cell follows the distribution $\Lap{\frac{1}{\epsilon}}$ and has a standard deviation of $\frac{\sqrt{2}}{\epsilon}$. Given an $m\times m$ grid, and a query that selects $r$ portion of the domain (where $r$ is the ratio of the area of the query rectangle to the area of the whole domain), about $rm^2$ cells are included in the query, and the total noise error thus has standard deviation of $\frac{\sqrt{2rm^2}}{\epsilon}=\frac{\sqrt{2r}m}{\epsilon}$.

The non-uniformity error is proportional to the number of data points in the cells that fall on the border of the query rectangle.  For a query that selects $r$ portion of the domain, it has four edges, whose lengths are proportional to $\sqrt{r}$ of the domain length; thus the query's border contains on the order of $\sqrt{r} m$ cells, which on average includes on the order of $\sqrt{r} m\times \frac{N}{m^2}=\frac{\sqrt{r} N}{m}$ data points.  Assuming that the non-uniformity error on average is some portion of the total density of the cells on the query border, then the non-uniformity error is $\frac{\sqrt{r} N}{c_0m}$ for some constant $c_0$.

To minimize the two errors' sum, $\frac{\sqrt{2r} m}{\epsilon} + \frac{\sqrt{r} N}{mc_0}$, we should set $m$ to $\sqrt{\frac{N\epsilon}{c}}$, where $c=\sqrt{2}c_0$.

%Since all the other cells are completely covered by the query, only the boundary cells contribute to this source of error. In the worst-case distribution, none of the points in these cells intersects with the query.
%Hence, the non-uniformity error is $c\frac{N}{n}$. To minimize the sum of both errors, $\frac{n}{\epsilon}+c\frac{N}{n}$, we should set $n$ to $\sqrt{N\epsilon/c}$.

Using Guideline~\ref{gl:size} requires knowing $N$, the number of data points.  Obtaining a noisy estimate of $N$ using a very small portion of the total privacy budget suffices.

The parameter $c$ depends on the uniformity of the dataset.  In the extreme case where the dataset is completely uniform, then the optimal grid size is $1\times 1$.  That is, the best method is to obtain as accurate a total count as possible, and then any query can be fairly accurately answered by computing what fraction of the region is covered by the query.  This corresponds to a large $c$.  When a dataset is highly non-uniform, then a smaller $c$ value is desirable.  In our experiments, we observe that setting $c= 10$ gives good results across datasets of different kinds.

\subsection{The Adaptive Grids Approach - \AG} \label{sec:approach:ag}

The main disadvantage of $\UG$ is that it treats all regions in the dataset equally. That is, both dense and sparse regions are partitioned in exactly the same way. This is not optimal. If a region has very few points, this method might result in \emph{over}-partitioning of the region, creating a set of cells with close to zero data points. This has the effect of increasing the noise error with little reduction in the non-uniformity error. On the other hand, if a region is very dense, this method might result in \emph{under}-partitioning of the region. As a result, the non-uniformity error would be quite large.

Ideally, when a region is dense, we want to use finer granularity partitioning, because the non-uniformity error in this region greatly outweighs that of noise error.  Similarly, when a region is sparse (having few data points), we want to use a more coarse grid there. Based on this observation, we propose an Adaptive Grids (\AG) approach.

The \AG approach works as follows. We first lay a coarse $m_1\times m_1$ grid over the data domain, creating $(m_1)^2$ first-level cells, and then we issue a count query for each cell using a privacy budget $\alpha\epsilon$, where $0<\alpha<1$.   For each cell, let $N'$ be the noisy count of the cell, \AG then partitions the cell using a grid size that is adaptively chosen based on $N'$, creating leaf cells.  The parameter $\alpha$ determines how to split the privacy budget between the two levels.

\mypara{Applying Constrained Inference.}  As discussed in Section~\ref{sec:related}, constrained inference has been developed in the context of one-dimensional histograms to improve hierarchical methods~\cite{HRMS10}.  The \AG method produces a 2-level hierarchy.  For each first-level cell, if it is further partitioned into a $m_2\times m_2$ grid, we can perform constrained inference.  %\todo{Perhaps include the detailed formula here.}

Let $v$ be the noisy count of a first-level cell, and let $u_{1,1}, \ldots, u_{m_2,m_2}$ be the noisy counts of the cells that $v$ is further partitioned into in the second level. One can then apply constrained inference as follows. First, one obtains a more accurate count $v'$ by taking the weighted average of $v$ and the sum of $u_{i,j}$ such that the standard deviation of the noise error at $v'$ is minimized.
%$$v' = \frac{\alpha^2 m_2^2}{\alpha^2(m^2+1) - 2\alpha +1} v + \frac{}{} \sum{u_{i,j}}$$
%$$v' = \frac{(1-\alpha)^2 m_2^2}{\alpha^2 + (1-\alpha)^2 m_2^2} v + \frac{\alpha^2}{\alpha^2 + (1-\alpha)^2 m_2^2} \sum{u_{i,j}}$$
$$v' = \frac{\alpha^2 m_2^2}{(1-\alpha)^2 + \alpha^2 m_2^2} v + \frac{(1-\alpha)^2}{(1-\alpha)^2 + \alpha^2 m_2^2} \sum{u_{i,j}}$$

This value is then propagated to the leaf nodes by distributing the difference among all nodes equally
$$u'_{i,j} = u_{i,j} + \left(v' - \sum{u_{i,j}}\right).$$
When $m_2=1$, the constrained inference step becomes issuing another query with budget $(1-\alpha)\epsilon$ and then computing a weighted average of the two noisy counts.
%One benefit of doing the above constrained inference is that when one includes significantly more than half of the leaf cells. \todo{complete.}

%The idea of constrained inference is that a large query would include many cells, thereby accumulating a large amount of noise error.  By grouping a number of adjacent cells together and issue another query for its sum, one could use this sum to more accurately answer queries that include all these cells.  The tradeoff is that one has to split the privacy budget between different levels, and the queries that include some, but not all the cells will have lower accuracy.  Our analysis in Section~\ref{sec:approach:idea} shows that we expect shallow hierarchies with large branching factors to perform better.

\mypara{Choosing Parameters for \AG.}
For the \AG method, we need to decide the formula to adaptively determine the grid size for each first-level cell.  We propose the following guideline.

\begin{guideline} \label{gl:size2}
Given a cell with a noisy count of $N'$, to minimize the errors, this cell should be partitioned into $m_2\times m_2$ cells, where $m_2$ is computed as follows:
$$\left\lceil \sqrt{\frac{N'(1-\alpha)\epsilon}{c_2}} \right\rceil,$$
where $(1-\alpha)\epsilon$ is the remaining privacy budget for obtaining noisy counts for leaf cells,  $c_2=c/2$, and $c$ is the same constant as in Guideline~\ref{gl:size}. 
%Our experimental results suggest that setting $c=10$ work well for the datasets we have experimented with.
\end{guideline}

The analysis to support this guideline is as follows.  When the first-level cell is further partitioned into $m_2\times m_2$ leaf cells, only queries whose borders go through this first-level cell will be affected.  These queries may include $0$, $1$, $2$, $\cdots$, up to $m_2-1$ rows (or columns) of leaf cells, and thus $0, m_2, 2m_2, \cdots, (m_2-1)m_2$ leaf cells.  When a query includes more than half of these leaf cells, constrained inference has the effect that the query is answered using the count obtained in the first level cell minus those leaf cells that are not included in the query.  Therefore, on average a query is answered using
 $$\frac{1}{m_2}\left(\sum_{i=0}^{m_2-1} \min(i, m_2-i)\right) m_2 \approx \frac{(m_2)^2}{4}$$
leaf cells, and the average noise error is on the order of $\sqrt{\frac{(m_2)^2}{4}}\frac{\sqrt{2}}{(1-\alpha)\epsilon}$.  The average non-uniformity error is about $\frac{N'}{c_0m_2}$; thereby to minimize their sum, we should choose $m_2$ to be about
$\sqrt {\frac{N' (1-\alpha)\epsilon}{\sqrt{2}c_0/2}}$.

The choice of $m_1$, the grid-size for the first level, is less critical than the choice of $m_2$.  When $m_1$ is larger, the average density of each cell is smaller, and the further partitioning step will partition each cell into fewer number of cells.  When $m_1$ is smaller, the further partitioning step will partition each cell into more cells.  In general, $m_1$ should be less than the grid size for \UG computed according to Guideline~\ref{gl:size}, since it will further partition each cell.  At the same time, $m_1$ should not be too small either.  We set
$$m_1 = \max\left(10,\frac{1}{4}\left\lceil \sqrt{\frac{N\epsilon}{c}}\right\rceil\right).$$
%We observe this works reasonably well in experiments.  

%When $\sqrt{\frac{N\epsilon}{c}}$, the predicated optimal \UG grid size is already small, e.g., $\leq 20$, then \AG is unlikely to outperform \UG, because there is little room to benefit from 

The choice of $\alpha$ also appears to be less critical.  Our experiments suggest that setting $\alpha$ to be in the range of $[0.2,0.6]$ results in similar accuracy.  We set $\alpha=0.5$.

%In order to minimize both sources of error, one can determine the grid size with which to partition by applying the same reasoning used for \UG above. That is, if $N'$ is the noisy count for some cell, one can partition that grid into $m_2 \times m_2$ cells, where $m_2$ is chosen based on $N'$. where $n_2 = \frac{\sqrt{N'\epsilon}}{c}$.

%In practice, we have found that setting $n_1 = ???$ produces the best results. Finally, one issues an additional count query for the new cells using privacy budget $\epsilon_2$. By setting $\epsilon_1 + \epsilon_2 = \epsilon$, one can achieve \eDP. In our experiments, we show the effects of various divisions of privacy budget.

\subsection{Comparing with Existing Approaches}\label{sec:approach:idea}

We now compare our proposed \UG and \AG with existing hierarchical methods, in terms of runtime efficiency, simplicity, and extensibility to higher dimensional datasets.

\mypara{Efficiency.}
The \UG and \AG methods are conceptually simple and easy to implement.  They also work well with very large datasets that cannot fit into memory.  \UG can be performed by a single scan of the data points.  For each data point, \UG just needs to increase the counter of the cell that the data point is in by 1.  \AG requires two passes over the dataset.  The first pass is similar to that of \UG. In the second pass, it first computes which first-level cell the data point is in, and then which leaf cell it is in. It then increases the corresponding counter.

We point out that another major benefit of \UG and \AG over recursive partition-based methods are their higher efficiency.  For all these methods, the running time is linear in the depth of the tree, as each level of the tree requires one pass over the dataset.  Existing recursive partitioning methods have much deeper trees (e.g., reaching 16 levels is common for 1 million data points).  Furthermore, these methods require expensive computation to choose the partition points.

\mypara{Effect of Dimensionality.}
Existing recursive partitioning approaches can be viewed as adapting the binary hierarchical method, which works well for 1-dimensional dataset, to the cases of 2 dimensions.  Some of these methods adapt quadtree methods, which can be viewed as extending 1-dimensional binary trees to 2 dimensions.  The emphasis is on how to perform the binary partition, e.g., using noisy mean, exponential method for finding the median, exponential method using non-uniformity measurement, etc.  The result is a deep tree.

We observe, however, while a binary hierarchical tree works well for the 1-dimensional case, their benefit for the 2-dimensional case is quite limited, and the benefit can only decrease with higher dimensionality.  When building a hierarchy, the interior of a query can be answered by higher-level nodes, but the borders of the query have to be answered using leaf nodes.  The higher the dimensionality, the larger the portion of the border region.

For example, for a 1-dimensional dataset with domain divided into $M$ cells, when one groups each $b$ adjacent cells into one larger cell, each larger cell is of size $\frac{b}{M}$ of the whole domain.  
Each query has $2$ border regions which need to be answered by leaf cells; each region is of size on the order of that of one larger cell, i.e., $\frac{b}{M}$ of the whole domain.
%can be answered using these large cells and at most $2b-2$ small cells.  The border of the query accounts for on the order of  $\frac{b}{M}$ of the total domain.
In the 2-dimensional case, with a $m\times m$ grid and a total of $M=m\times m$ cells, if one groups $b=\sqrt{b}\times \sqrt{b}$ adjacent cells together, then a query's border, which needs to be answered by leaf nodes, has $4$ sides, and each side is of size on the order of $\frac{\sqrt{b}}{\sqrt{M}}$ of the whole domain.  Note that $4\frac{\sqrt{b}}{\sqrt{M}}$ is much larger than $2\frac{b}{M}$, since $M$ is always much larger than $b$.  For example, when $M=10,000$ and $b=4$, $4\frac{\sqrt{b}}{\sqrt{M}}=0.08$, and $2\frac{b}{M}=0.0008$.

%on average case cuts through $4 \times m/{2d}$ big cells, which accounts for on the order of $2\sqrt{b}/\sqrt{M}$ of the whole domain,
Therefore, in 2-dimensional case, one benefits much less from a hierarchy, which provides less accurate counts for the leaf cells.  This effect keeps growing with dimensionality.  For $d$ dimensions, the border of a query has $2d$ hyperplanes, each of size on the order of $\frac{\sqrt[d]{b}}{\sqrt[d]{M}}$.  In our experiments, we have observed some small benefits for using hierarchies, which we conjecture will disappear with $3$ or higher dimensional cases.

This analysis suggests that our approach of starting from the Uniform Grid method and trying to improve upon this method is more promising than trying to improve a hierarchical tree based method.  When focusing on Uniform Grid, the emphasis in designing the algorithm shifts from choosing the axis for partitioning to choosing the partition granularity.  When one partitions a cell into two sub-cells, the question of how to perform the partitioning depending on the data in the cell seems important and may affect the performance; and thus one may want to use part of the privacy budget to figure out what the best partitioning point is.  On the other hand, when one needs to partition a cell into, e.g., $8\times 8$, sub-cells in a differentially private way, it appears that the only feasible solution is to do equi-width partition.  Hence the only parameter of interest is what is the grid size.

%In short the higher the number of dimensions, the more volume the boundary of a shape has relative to the interior of the shape (assuming that the boundary has a certain width).  Thus, when one goes beyond 1-dimensional dataset, one should replace recursive binary partitioning with methods based on \UG.

\comment{
\subsection{The Differentiated Adaptive Grids Approach - \DAG} \label{sec:approach:dag}
%Our method uses three levels of granularities.  Main level.  Upper level.  Lower level.

Further improvement over the \AG method can be obtained by observing that the privacy budget is not always fully utilized.  After the first level, there are two kinds of cells.  For those cells that have higher densities and will be further partitioned, the remaining privacy budget is used for cells of the next level.  For those cells that have low density and will not be further partitioned, however, the privacy budget is somewhat wasted.  One could issue another count count query for these cells and use the average of the old noisy count and the new noisy count to obtain a more accurate count.  (When $\alpha=0.5$, this provides a $1\sqrt{2}$ reduction in the standard error.  A more effective use of the privacy budget is to construct another level above these nodes and use the privacy budget for .

Instead, we propose a differentiated approach which treats these cells differently. We call this the Differentiated Adaptive Grids approach (\DAG), and it works as follows. We perform the first level partitioning similar to \AG. Then, if a region is dense, we follow the \AG approach. In addition, we group all sparse regions together and issue a coarser granularity  count for their union. We then perform constrained inference as described above.

In essence, this can be viewed as performing partitioning at three different granularity. However, each individal cell participates in two levels only. Thus if $\epsilon$ is divided among the two partitioning steps, then the resulting algorithm would satisfy $\eDP$. \todo{equal division of privacy budget because...}

Now we can choose $m_1$ to be finer grained than the case of $\AG$.

$$m_1 = \max\left(10,\frac{1}{2}\sqrt{\frac{N\epsilon}{c}}\right)$$

} 

%\section{Relationship with Other Methods}
%\todo{In this section, we discuss why the one-dimensional methods would not result in significant improvements.}

\section{Experimental Results} \label{sec:results}
%!TEX root = ./md_hist.tex

\subsection{Methodology}

We have conducted extensive experiments using four real datasets, to compare the accuracy of different methods and to validate our analysis of the choice of parameters.

\begin{figure*}
		\subfigure[The road dataset]{\label{fig:data1}
\includegraphics[width=0.24\textwidth]{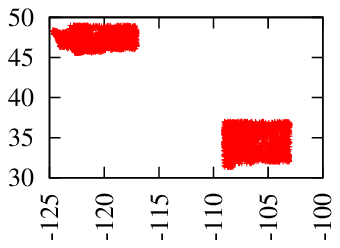}}%
	\hfill
	\subfigure[The checkin dataset]{\label{fig:data2}
\includegraphics[width=0.24\textwidth]{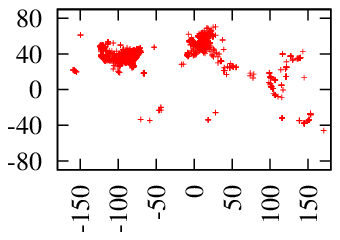}}%
    \hfill
	\subfigure[The landmark dataset]{\label{fig:data3}
  \includegraphics[width=0.24\textwidth]{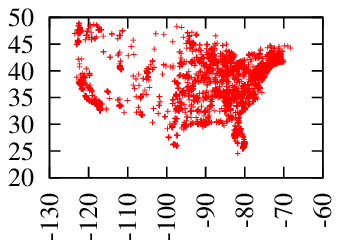}}%
	\hfill
	\subfigure[The storage dataset]{\label{fig:data3b}
 \includegraphics[width=0.24\textwidth]{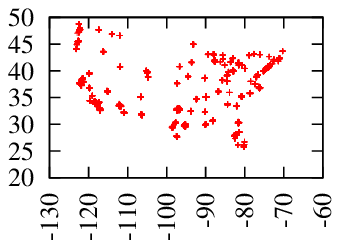}}
	\caption{Illustration of datasets.} \label{fig:datasets}
\end{figure*}

\mypara{Datasets.}
We illustrate these datasets by plotting the data points directly in Figure~\ref{fig:datasets}. We also present the parameters for these datasets in Table~\ref{table:datasets}.

The first dataset (which we call the ``road'' dataset) includes the GPS coordinates of road
intersections in the states of Washington and New Mexico, obtained from 2006 TIGER/Line from US Census. This is the dataset that was used in \cite{CPS+12} for experimental evaluations. There are about 1.6M data points.
As illustrated in Figure~\ref{fig:data1}, the distribution of the data points is quite unusual.  There are large blank areas with two dense regions (corresponding to the two states).
%If we measure the accuracy of query answering, the result will mainly based on the way handling those blank region, which is not what we want. Therefore, we introduced 3 other real datasets to experiment on.

The second dataset is derived from the checkin dataset~\footnote{http://snap.stanford.edu/data/loc-gowalla.html} from the Gowalla location-based social networking website, where users share their locations by checking-in. This dataset records time and location information of check-ins made by users over the period of Feb.~2009 - Oct.~2010. We only use the location information.  There are about 6.4M data points.  The large size of the dataset makes it infeasible to run the implementation of KD-tree based methods obtained from authors of~\cite{CPS+12} due to memory constraints. % on the laptop we run experiments.
We thus sampled 1M data points from this dataset, and we call this the ``checkin'' dataset.  As illustrated in Figure~\ref{fig:data2}, the shape vaguely resembles a world map, but with the more developed and/or populous areas better represented than other areas.

We obtained both the third dataset (``landmark'' dataset) and the fourth dataset (``storage'' datset) from infochimps.  The landmark dataset \footnote{http://www.infochimps.com/datasets/storage-facilities-by-landmarks} consists of locations of landmarks in the 48 continental states in the United State.  The listed landmarks range from schools and post offices to shopping centers, correctional facilities, and train stations from the 2010 Census TIGER point landmarks.  There are over 870k data points.  As illustrated in Figure~\ref{fig:data2}, the dataset appears to match the population distribution in US.

The storage dataset \footnote{http://www.infochimps.com/datasets/storage-facilities-by-neighborhood--2} includes US storage facility locations. Included are national chain storage facilities, as well as locally owned and operated facilities. This is a small dataset, consisting about 9000 data points.  We chose to use this dataset to analyze whether our analysis and guideline in Section~\ref{sec:approach} holds for both large and small datasets.

\mypara{Absolute and Relative Error.}
Following~\cite{CPS+12}, we primarily consider the relative error, defined as follows:
For a query $r$,
we use $A(r)$ to denote the correct answer to $r$.  For a method $\MM$ and a query $r$, we use $Q_\MM(r)$ to denote the answer to the query $r$ when using the histogram constructed by method $\MM$ to answer the query $r$, then the relative error is defined as
$$RE_{\MM}(r) = \frac{|Q_\MM(r)-A(r)|}{\max\{A(r), \rho\}}$$
where we set $\rho$ to be $0.001*|D|$, where $D$ is the total number of data points in $D$.  This avoids dividing by 0 when $A(r) = 0$.

$RE_{\MM}(r)$ is likely to be largest when the query $r$ is mid-size.  When the range of a query, $r$, is large, $RE_{\MM}(r)$ is likely to be small since $A(r)$ is likely to be large.  On the other hand, when the range of $r$ is small, the absolute error $|Q_\MM(r)-A(r)|$ is likely to be small.

While we primarily use relative error, we also use absolute error in the final comparison.

%We aim at minimizing relative error of query answering. Relative error is defined as $rel\_err = \frac{|noisy\_count - actual\_count|}{d}$ and $d = min(actual\_count, 0.001N)$. We do not directly use $actual\_count$ as denominator because to $actual\_count$ may be too small or even can be zero. In our experiment, we will present relative error at 25 percentile, 25 percentile, 75 percentile and 95 percentile and also the average of relative error. We introduced candlestick to show these 5 values. In the following figures, we will use the upper bound of the line to represent 95 percentile relative error, use lower bound of the line to represent 25 percentile relative error, use upper bound of bar to represent 75 percentile relative error, use lower bound of bar to represent 50 percentile relative error, and use a black line to represent mean of relative error.

\mypara{Understanding the Figures.}
We use two $\epsilon$ values, $\epsilon=0.1$ and $\epsilon=1$.  For each algorithm, we use $6$ query sizes, with $q_1$ being the smallest, each $q_{i+1}$ doubles both the $x$ range and $y$ range of $q_{i}$, thereby quadrupling the query area, and $q_6$, the largest query size covering between $1/4$ and $1/2$ of the whole space.  The query sizes we have used are given in Table~\ref{table:datasets}.

For each query size, we randomly generate $200$ queries, and compute the errors in answering them.  We use two kinds of graphs.  To illustrate the results across different query sizes, we use line graphs to plot the arithmetic mean of the relative error for each query size.  To provide a clearer comparison among different algorithms, we use candlesticks to plot the profile of relative errors for all query sizes.  Each candlestick provides 5 pieces of information: the 25 percentile (the bottom of candlestick), the median (the bottom of the box), the 75 percentile (the top of the box), the 95 percentile (the top of the candlestick), and the arithmetic mean (the black bar).  
 %We use 95 percentile instead of max, because the max tends to be much larger than the other values.  
We pay the most attention to the arithmetic mean.

\begin{table}
\begin{center}
\begin{tabular}{|l|l|} \hline
K$_\mathrm{st}$ & KD-standard
 \\ \hline
K$_\mathrm{hy}$ & KD-hybrid
 \\ \hline
U$_m$ & \UG with $m\times m$ grid
 \\ \hline
W$_m$ & Privlet with $m\times m$ grid
 \\ \hline
H$_{b,d}$ & Hierarchy with $d$ levels and $b\times b$ branching
 \\ \hline
A$_{m_1,c_2}$ & \AG with $m_1\times m_1$ grid and the given $c_2$ value
 \\ \hline
\end{tabular}
\end{center}
\caption{Notation for Algorithms.}\label{table:algorithms}
\end{table}

\mypara{Algorithm Notation.}  The notation for the algorithms we use in our experiments are given in Table~\ref{table:algorithms}.  
 %H$_{b,d}$ is only used in Figure~\ref{fig:hie_flat}, where they use a $360\times 360$ base grid.  
The \AG method is denoted by A$_{m_1,c_2}$, which first lays a $m_1\times m_1$ grid, then uses $\alpha\epsilon$ to issue count query for each cell. In addition, it partitions each cell with noisy count $N'$ into $m_2\times m_2$ grid, with $m_2=\left\lceil \sqrt \frac{N'(1-\alpha)\epsilon}{c_2} \right\rceil$.  Unless explicitly noted, $\alpha$ is set to be $0.5$.

\begin{table*}
\begin{center}
\begin{tabular}{|l|c|c|c|c|c|c|c|c|c|c|}
 \hline
 \multirow{2}{*}{dataset} & \multirow{2}{*}{\# of points} & \multirow{2}{*}{domain size} & \multirow{2}{*}{size of $q_6$}  & \multirow{2}{*}{size of $q_1$}
 %{dataset} & {\# of points} & {domain size} & {size of $q_6$}  & {size of $q_1$}
 & \multicolumn{3}{|c|}{best grid size $\epsilon=1$}
 & \multicolumn{3}{|c|}{best grid size $\epsilon=0.1$}
 \\
 & & & & & \UG sugg. & \UG actual & \AG actual &  \UG sugg. & \UG actual & \AG actual
 \\ \hline
road & 1.6M & $25\times 20$ & $16\times 16$ & $0.5 \times 0.5$ & 400 & 96-192 & 32-48 & 126 & 48-128 & 10-32
 \\ \hline
checkin & 1M & $360 \times 150$ & $192\times 96$ & $6\times 3$ & 316 & 192-384 & 48-96 & 100 & 64-128 & 16-48
 \\ \hline
landmark & 0.9M & $60 \times 40$ &  $40 \times 20$ & $1.25 \times 0.625$ & 300 & 256-512 & 64-128 &95 & 64-128 & 32-64
 \\ \hline
storage & 9K &$60 \times 40$ & $40 \times 20$& $1.25 \times 0.625$ &30& 32-64 & 12-32 & 10 & 10-32 & 10-16

 \\ \hline
\end{tabular}
\end{center}
 \caption{Experimental Information About Datasets. \label{table:datasets}} Columns are dataset name, number of data points, domain size, the largest query size $q_6$ in experiments, the smallest query size $q_1$, and three grid sizes each for $\epsilon=1$ and $\epsilon=0.1$, including the grid size suggested by Guideline~\ref{gl:size}, the range of grid sizes that perform the best in the experiments with \UG, and the range of best-performing sizes for \AG.
\end{table*}

\subsection{Comparing KD-Tree with \UG}

\begin{figure*}
		\subfigure[road, $\epsilon = 0.1$]{\label{fig:kd_1_1} \includegraphics[width=0.24\textwidth]{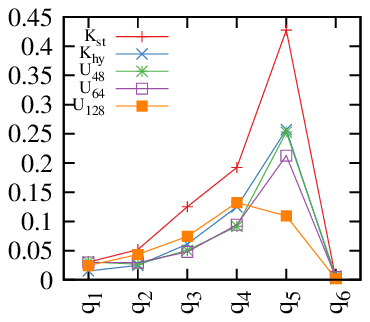}}%
	\hfill
	\subfigure[road, $\epsilon = 0.1$]{\label{fig:kd_1_2}
\includegraphics[width=0.24\textwidth]{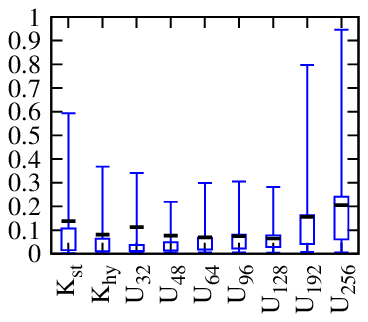}}%
    \hfill
	\subfigure[road, $\epsilon = 1$]{\label{fig:kd_1_3}
  \includegraphics[width=0.24\textwidth]{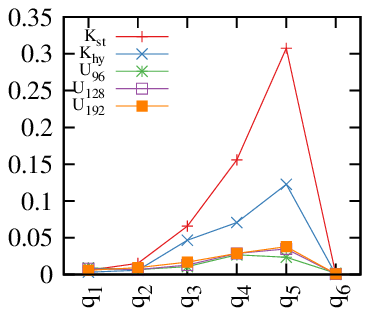}}%
	\hfill
	\subfigure[road, $\epsilon = 1$]{\label{fig:kd_1_4}
 \includegraphics[width=0.24\textwidth]{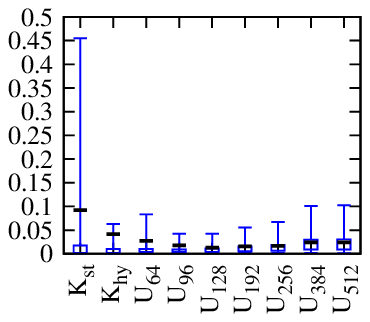}}

\subfigure[checkin, $\epsilon = 0.1$]{\label{fig:kd_2_1} \includegraphics[width=0.24\textwidth]{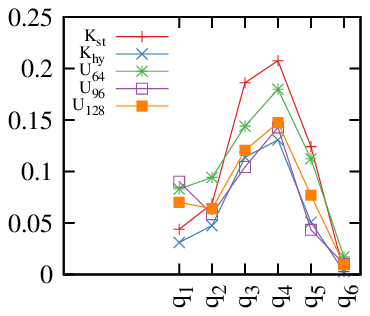}}%
	\hfill
	\subfigure[checkin, $\epsilon = 0.1$]{\label{fig:kd_2_2}
\includegraphics[width=0.24\textwidth]{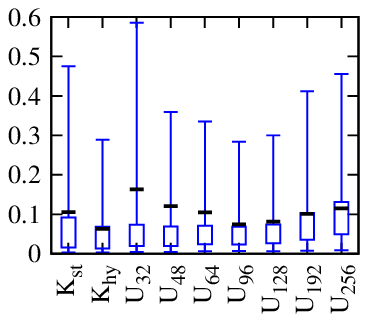}}%
    \hfill
	\subfigure[checkin, $\epsilon = 1$]{\label{fig:kd_2_3}
  \includegraphics[width=0.24\textwidth]{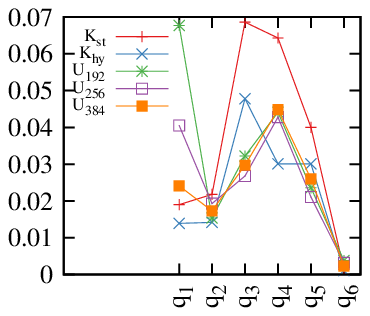}}%
	\hfill
	\subfigure[checkin, $\epsilon = 1$]{\label{fig:kd_2_4}
 \includegraphics[width=0.24\textwidth]{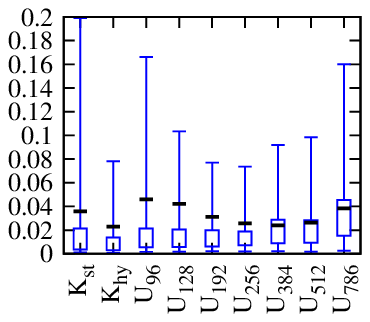}}

 \subfigure[landmark, $\epsilon = 0.1$]{\label{fig:kd_3_1} \includegraphics[width=0.24\textwidth]{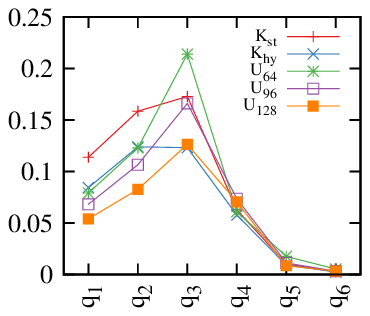}}%
	\hfill
	\subfigure[landmark, $\epsilon = 0.1$]{\label{fig:kd_3_2}
\includegraphics[width=0.24\textwidth]{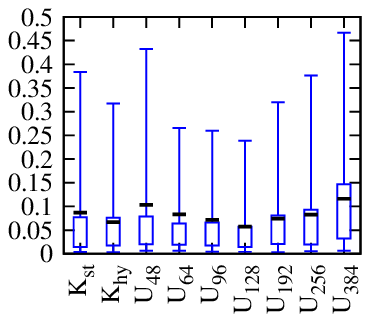}}%
    \hfill
	\subfigure[landmark, $\epsilon = 1$]{\label{fig:kd_3_3}
  \includegraphics[width=0.24\textwidth]{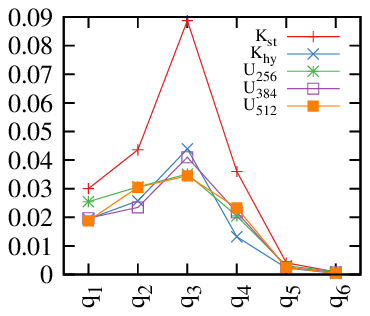}}%
	\hfill
	\subfigure[landmark, $\epsilon = 1$]{\label{fig:kd_3_4}
 \includegraphics[width=0.24\textwidth]{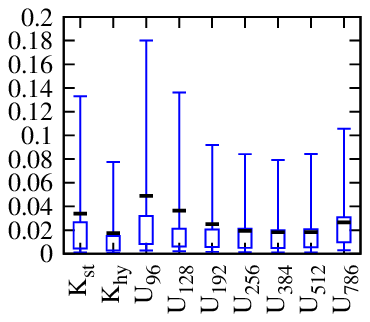}}

  \subfigure[storage, $\epsilon = 0.1$]{\label{fig:kd_4_1} \includegraphics[width=0.24\textwidth]{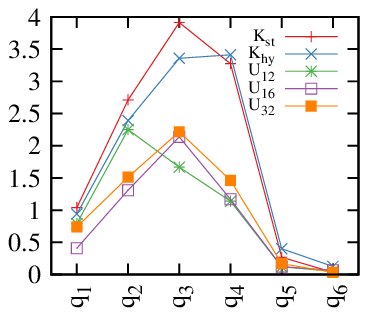}}%
	\hfill
  \subfigure[storage, $\epsilon = 0.1$]{\label{fig:kd_4_2}
\includegraphics[width=0.24\textwidth]{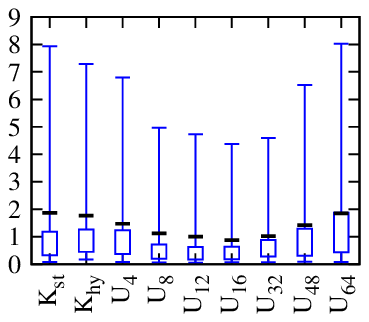}}%
    \hfill
	\subfigure[storage, $\epsilon = 1$]{\label{fig:kd_4_3}
  \includegraphics[width=0.24\textwidth]{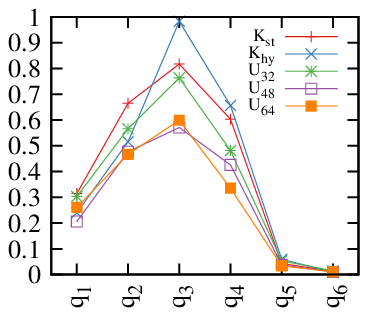}}%
	\hfill
	\subfigure[storage, $\epsilon = 1$]{\label{fig:kd_4_4}
 \includegraphics[width=0.24\textwidth]{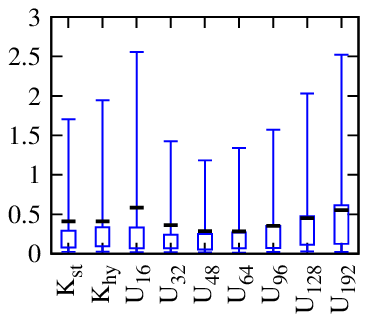}}

	\caption{Comparing KD-standard, KD-hybrid and \UG with different sizes. } \label{fig:kd1}
\end{figure*}

In the first set of experiments, we compare KD-standard, KD-hybrid with \UG with different grid sizes, and we identify the best performing grid size for \UG.  The results are presented in Figure~\ref{fig:kd1}.

\mypara{Analysis of Results.}
We can observe that generally the relative errors are maximized at queries of the middle sizes.  More specifically, the maximizing points are $q_5$ for the road dataset, $q_4$ for the checkin dataset, and $q_3$ for landmark and storage.  We believe this is due to the existence of large blank areas in the road dataset and the checkin dataset.  The large blank areas cause large queries to have low true count, which cause large relative errors due to the large noise error for large queries.

We can see that when varying the grid size for the \UG method, there exist a range of sizes where the methods perform the best.  Larger or smaller sizes tend to perform worse.  When leaving the optimal range, the error steadily increases.  This suggests that choosing a good grid size is important.

The ranges for the experimentally observed optimal grid sizes are give in Table~\ref{table:datasets}.  We can see that Guideline~\ref{gl:size} works remarkably well.  The predicted best \UG size generally lie within the range of the sizes that experimentally perform the best, and often fall in the middle of the range.  In two cases, the predicted size lies outside the observed optimal range.
For the storage dataset with $\epsilon=1$, the predicted \UG size is $30$, which is quite close to $32-64$, the range of the sizes observed to have lowest.  Only on the road dataset (which has unusually high uniformity) at $\epsilon=1$, our prediction (400) lies outside the observed optimal range (96-192).  However, we observe that even though the high uniformity calls for a smaller optimal grid size, the performance at grid sizes 384 and 512 is quite reasonable; indeed, the average relative error in both cases are still lower than that of KD-hybrid.  Jumping ahead, in Figure~\ref{fig:ag1_2_2} we will see that U$_{400}$ significantly outperforms U$_{96}$ in terms of absolute error, further validating Guideline~\ref{gl:size}.

We can also see that the KD-hybrid method performs worse than the best \UG method on the road dataset and the storage dataset, and is very close to the best \UG method on the other two datasets.

\begin{figure*}
\subfigure[checkin, $\epsilon = 0.1$]{\label{fig:hie_2_1} \includegraphics[width=0.24\textwidth]{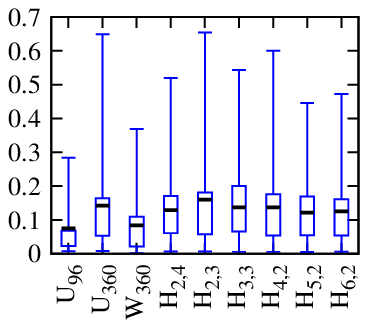}}%
	\hfill
	\subfigure[checkin, $\epsilon = 1$]{\label{fig:hie_2_2}
\includegraphics[width=0.24\textwidth]{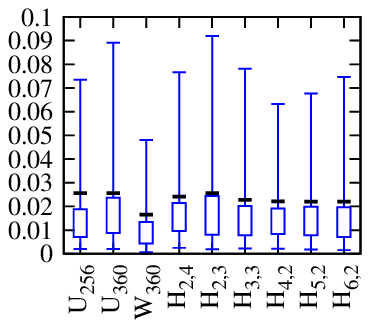}}%
    \hfill
	\subfigure[landmark, $\epsilon = 0.1$]{\label{fig:hie_3_1}
  \includegraphics[width=0.24\textwidth]{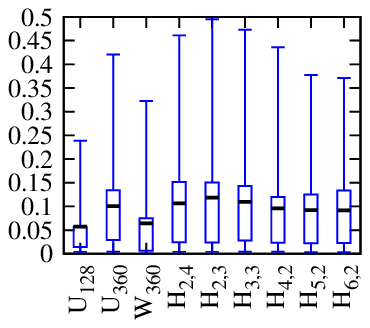}}%
	\hfill
	\subfigure[landmark, $\epsilon = 1$]{\label{fig:hie_3_2}
 \includegraphics[width=0.24\textwidth]{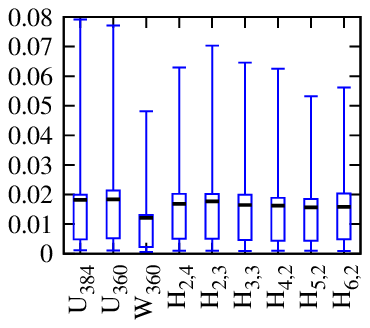}}

	\caption{Analyzing the effect of Hierarchies.  In each figure, the first algorithm is \UG with experimentally observed best grid size, the second is \UG with 360, and the rest build hierarchies on top of a $360\times 360$ grid. $H_{b,d}$ means build a hierarchy with a $b\times b$ branching factor and depth $d$.} \label{fig:hie_flat}
\end{figure*}

\mypara{Effect of Adding Hierarchies.}
In Figure~\ref{fig:hie_flat}, we evaluate the effect of adding hierarchies to \UG to improve its accuracy.  Our goal is to understand whether adding hierarchies of different branching factor to \UG would result in better accuracy.  Here we present only results for the checkin and the landmark dataset, because the road dataset is unusual, and the storage dataset is too small to benefit from a hierarchy.

We include results for the \UG method with the lowest observed relative error, the \UG method with $m=360$, which is close to the size suggested by Guideline~\ref{gl:size} and is multiples of many numbers, facilitating experiments with different branching factors.  We also include results for W$_{360}$, which applies the Privlet~\cite{XWG11} method, described in Section~\ref{sec:related}, to leaf cells from a $360\times 360$ grid.  We consider hierarchical methods with branching factors ranging from $2\times 2$ to $b\times b$.  We also vary depths of the tree for the branching factor $2\times 2$.  That is H$_{2,3}$ uses $3$ levels, with sizes at $360,180,90$.

From the results we observe that while adding hierarchies can somewhat improve the accuracy, the benefit is quite small.  In Section~\ref{sec:approach:idea}, we have analyzed the reason for this. 
Applying Privlet, however, results in clear, if not significant, accuracy improvements.  
This can be attributed to the noise reduction effect that the Privlet method has over general hierarchical methods.
%Since the wavelet transform is applied twice to each cell in the grid, the sensitivity of the differentially private queries are smaller, and therefore, the noise is smaller.
Jumping slightly ahead to Figure~\ref{fig:compare}, we observe, however, applying Privlet to smaller grid sizes (e.g., $\leq 128$) tends to be worse the \UG.  

\subsection{Evaluating Adaptive Grids}

\begin{figure*}
\subfigure[checkin, $\epsilon=0.1$]{ \includegraphics[width=0.24\textwidth]{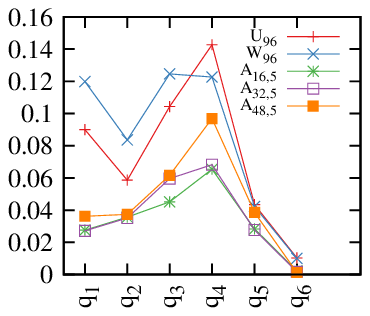}}%
\hfill
\subfigure[checkin, $\epsilon=0.1$, vary $m_1$,\newline suggested $m_1=25$]{ \includegraphics[width=0.24\textwidth]{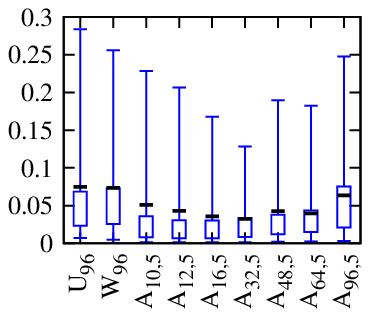}}%
\hfill
\subfigure[checkin, $\epsilon=0.1$, fix $m_1=16$, \newline vary $\alpha$ and $c_2$]{ \includegraphics[width=0.24\textwidth]{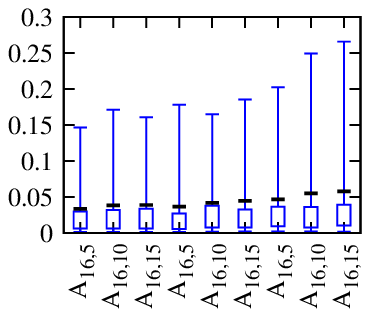}}%
\hfill
\subfigure[checkin, $\epsilon=0.1$, fix $m_1=32$, \newline vary $\alpha$ and $c_2$]{ \includegraphics[width=0.24\textwidth]{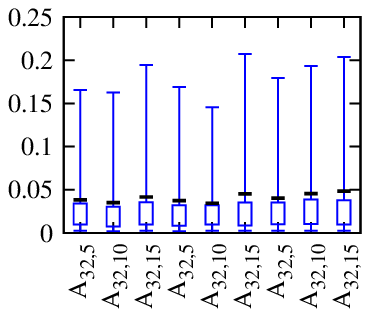}}%

\subfigure[checkin, $\epsilon=1$]{ \includegraphics[width=0.24\textwidth]{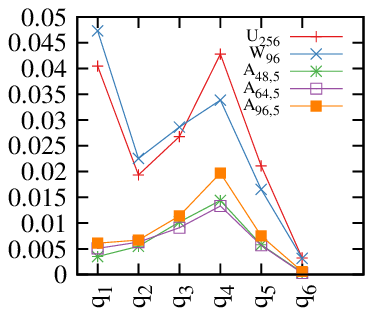}}%
\hfill
\subfigure[checkin, $\epsilon=1$, vary $m_1$, \newline suggested $m_1=79$]{ \includegraphics[width=0.24\textwidth]{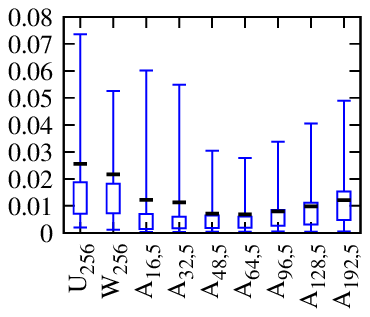}}%
\hfill
\subfigure[checkin, $\epsilon=1$, fix $m_1=48$, \newline vary $\alpha$ and $c_2$]{ \includegraphics[width=0.24\textwidth]{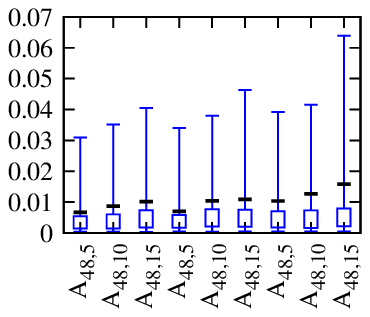}}%
\hfill
\subfigure[checkin, $\epsilon=1$, fix $m_1=64$, \newline vary $\alpha$ and $c_2$]{ \includegraphics[width=0.24\textwidth]{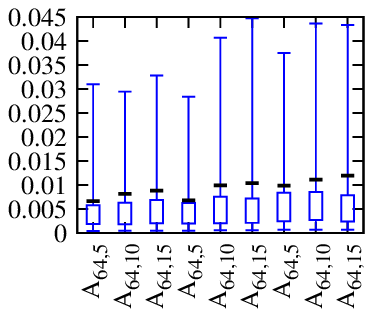}}%
	
	\subfigure[landmark, $\epsilon=0.1$]{
\includegraphics[width=0.24\textwidth]{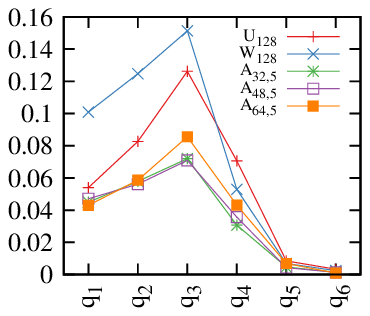}}%
\hfill
\subfigure[landmark, $\epsilon=0.1$, vary $m_1$, \newline suggested $m_1=24$]{ \includegraphics[width=0.24\textwidth]{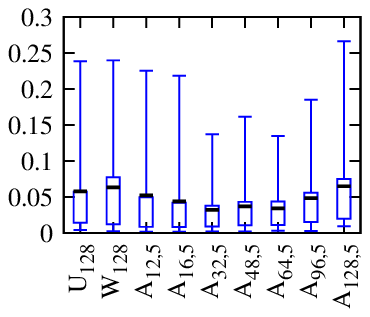}}%
\hfill
\subfigure[landmark, $\epsilon=0.1$, fix $m_1=32$, \newline vary $\alpha$ and $c_2$]{ \includegraphics[width=0.24\textwidth]{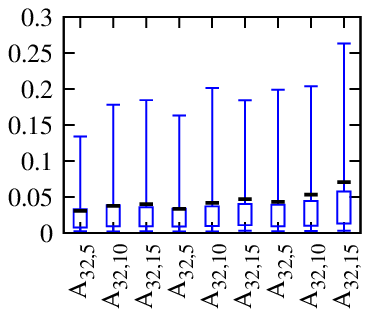}}%
\hfill
\subfigure[landmark, $\epsilon=0.1$, fix $m_1=64$, \newline vary $\alpha$ and $c_2$]{ \includegraphics[width=0.24\textwidth]{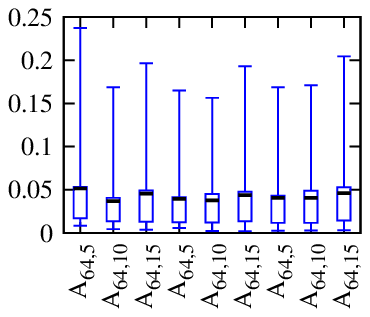}}%

\subfigure[landmark, $\epsilon=1$]{ \includegraphics[width=0.24\textwidth]{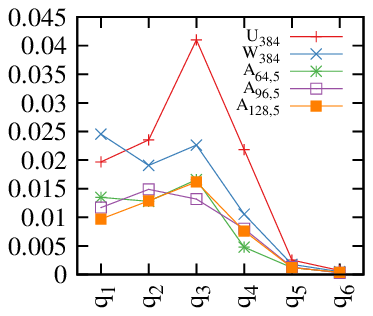}}%
\hfill
\subfigure[landmark, $\epsilon=1$, vary $m_1$, \newline suggested $m_1=75$]{ \includegraphics[width=0.24\textwidth]{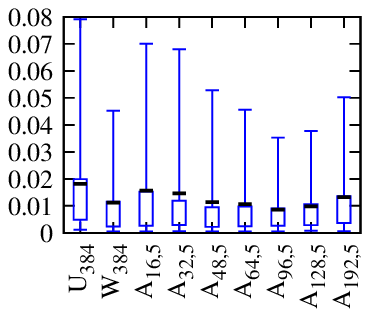}}%
\hfill
\subfigure[landmark, $\epsilon=1$, fix $m_1=48$, \newline vary $\alpha$ and $c_2$]{ \includegraphics[width=0.24\textwidth]{fig/ag0_2_3.eps}}%
\hfill
\subfigure[landmark, $\epsilon=1$, fix $m_1=64$, \newline vary $\alpha$ and $c_2$]{ \includegraphics[width=0.24\textwidth]{fig/ag0_2_4.eps}}%

	\caption{Varying the parameters for the \AG method. For figures in the first and second columns, $\alpha=0.5$.  For the third and fourth columns, each figure has $9$ candlesticks, the left three use $\alpha=0.25$, the middle three use $\alpha=0.5$, and the right three use $\alpha=0.75$.} \label{fig:ag0_param}
\end{figure*}

Figure~\ref{fig:ag0_param} presents experimental results on the effect of choosing different parameters for the \AG method.  We use checkin and landmark datasets. 
 %A$_{m_1,c_2}$ denotes the algorithm to first lay an $m_1\times m_1$ grid, uses $\alpha\epsilon$ to issue count query for each cell, then partitions each cell with noisy count $N'$ into $m_2\times m_2$ grid, where $m_2=\lceil \sqrt \frac{N'(1-\alpha)\epsilon}{c_2} \sqrt\rceil$. For figures in the first and second columns, $\alpha=0.5$.
The first column shows comparison of the three \AG methods with best performing grid sizes with the best-performing \UG method and the Privlet method with the same grid size.  We show results for different query sizes.  We see that the \AG methods outperform \UG and Privlet across all query sizes.

The second column shows the effect of varying $m_1$, the first-level grid size of the \AG method.  We see that while the \AG method like the \UG method is affected by $m_1$, it is less sensitive to $m_1$ and provides good performance for a wider range of $m_1$, and that the $m_1$ suggested by Guideline~\ref{gl:size2} is either at or close to the optimal size.

The third and the fourth columns explore the effect of varying $\alpha$ and $c_2$.  In each figure, there are $9$ candlesticks, divided into $3$ groups of $3$ each. The left group uses $\alpha=0.25$, the middle group uses $\alpha=0.5$, and the right group uses $\alpha=0.75$.  Within each group, we vary the value of $c_2$.  As can be seen, setting $c_2=c/2 =5$ as suggested significantly outperforms larger values of $c_2$, namely $10$ and $15$.  We have also conducted experiments with $c_2$ values from $3$ to $9$, and the results (which we did not include in the paper for space limitation) show that setting $c_2$ in the range of $3$ to $7$ result in almost the same accuracy.  The effect of varying $\alpha$ can be seen by comparing the left group of $3$, with the middle group, and the right group.  We observe that setting $\alpha=0.75$ performs worse than the other $\alpha$ values.  Setting $\alpha=0.25$ and $\alpha=0.5$ give very similar results, perhaps with $\alpha=0.25$ slightly better.  We have also experimented with setting $\alpha$ from $0.1$ to $0.9$, with increment of $0.1$.  The results suggest that setting $\alpha$ in the range of $0.2$ to $0.6$ give very similar results.  We use $\alpha=0.5$ as the default value.

\subsection{Final Comparison}

\begin{figure*}
\subfigure[road, $\epsilon=0.1$]{ \includegraphics[width=0.24\textwidth]{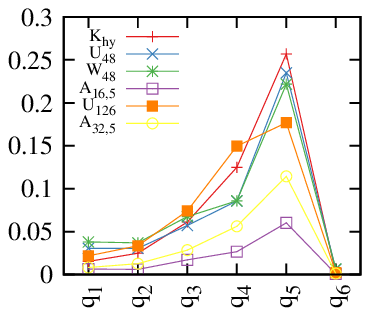}}%
\hfill
\subfigure[road, $\epsilon=0.1$]{ \includegraphics[width=0.24\textwidth]{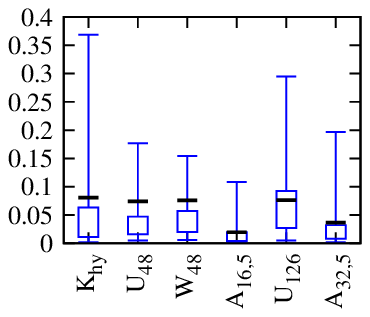}}%
\hfill
\subfigure[road, $\epsilon=1$]{ \includegraphics[width=0.24\textwidth]{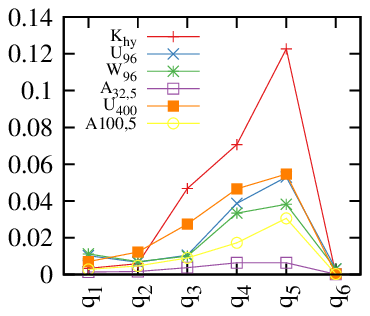}}%
\hfill
\subfigure[road, $\epsilon=1$]{ \includegraphics[width=0.24\textwidth]{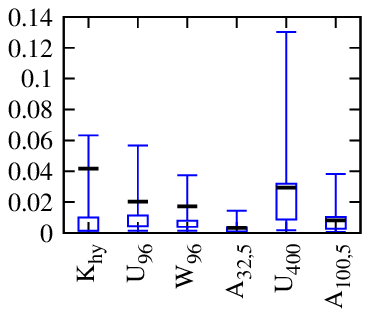}}%

\subfigure[checkin, $\epsilon=0.1$]{ \includegraphics[width=0.24\textwidth]{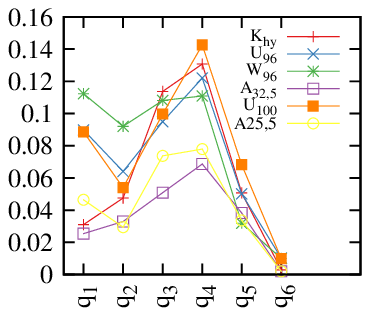}}%
\hfill
\subfigure[checkin, $\epsilon=0.1$]{ \includegraphics[width=0.24\textwidth]{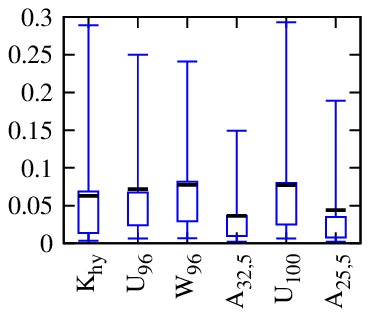}}%
\hfill
\subfigure[checkin, $\epsilon=1$]{ \includegraphics[width=0.24\textwidth]{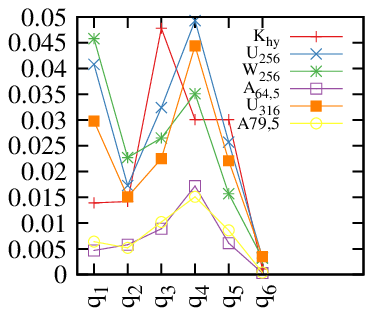}}%
\hfill
\subfigure[checkin, $\epsilon=1$]{ \includegraphics[width=0.24\textwidth]{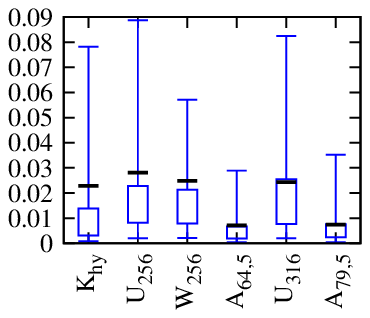}}%
	
	\subfigure[landmark, $\epsilon=0.1$]{
\includegraphics[width=0.24\textwidth]{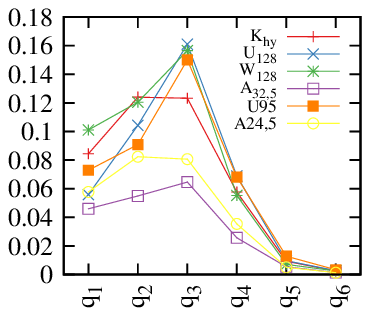}}%
\hfill
\subfigure[landmark, $\epsilon=0.1$]{ \includegraphics[width=0.24\textwidth]{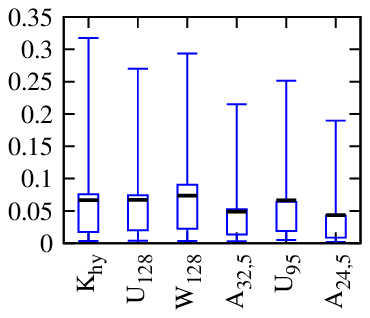}}%
\hfill
\subfigure[landmark, $\epsilon=1$]{ \includegraphics[width=0.24\textwidth]{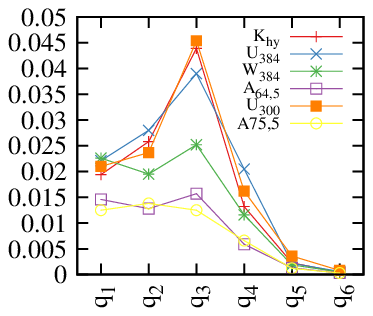}}%
\hfill
\subfigure[landmark, $\epsilon=1$]{ \includegraphics[width=0.24\textwidth]{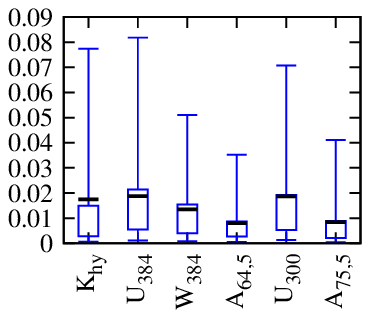}}%

	\subfigure[storage, $\epsilon=0.1$]{
\includegraphics[width=0.24\textwidth]{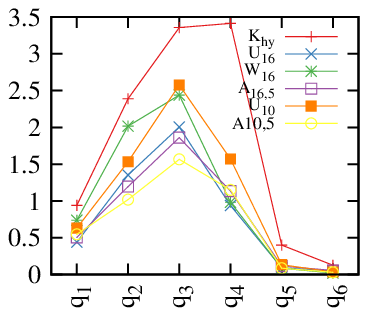}}%
\hfill
\subfigure[storage, $\epsilon=0.1$]{ \includegraphics[width=0.24\textwidth]{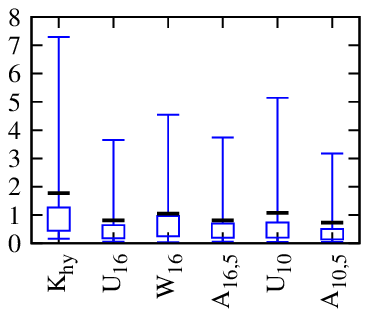}}%
\hfill
\subfigure[storage, $\epsilon=1$]{ \includegraphics[width=0.24\textwidth]{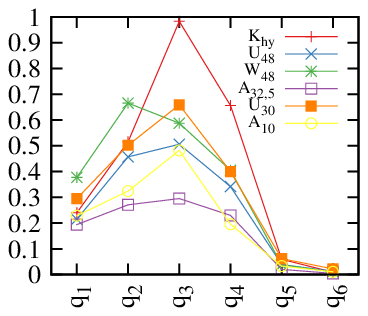}}%
\hfill
\subfigure[storage, $\epsilon=1$]{ \includegraphics[width=0.24\textwidth]{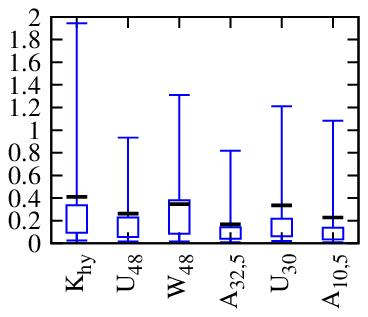}}%

	\caption{Comparing, from left to right, KD-hybrid, \UG with size giving lowest observed relative error, Privlet on this grid size, \AG with $m_1$ giving lowest observed relative error, \UG with suggested size, \AG with suggested size.} \label{fig:compare}
\end{figure*}

\begin{figure*}
\subfigure[road, $\epsilon=0.1$]{\label{fig:ag1_2_1} \includegraphics[width=0.24\textwidth]{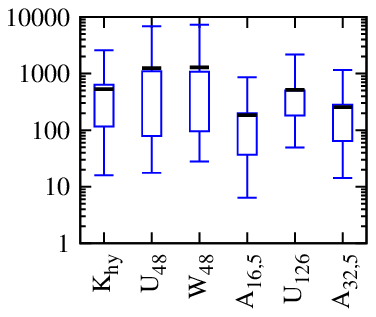}}%
\hfill
\subfigure[road, $\epsilon=1$]{\label{fig:ag1_2_2} \includegraphics[width=0.24\textwidth]{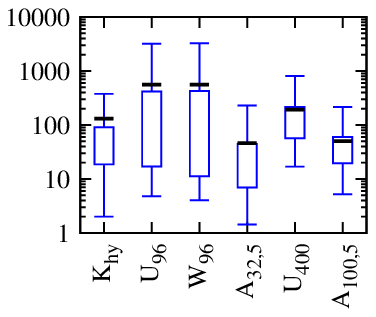}}%
\hfill
\subfigure[checkin, $\epsilon=0.1$]{\label{fig:ag1_2_3} \includegraphics[width=0.24\textwidth]{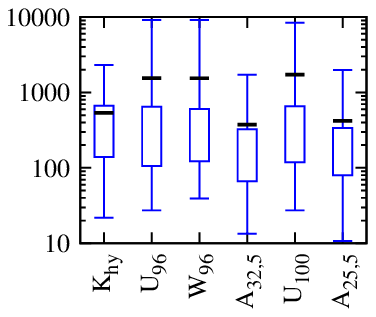}}%
\hfill
\subfigure[checkin, $\epsilon=0.1$]{\label{fig:ag1_2_4} \includegraphics[width=0.24\textwidth]{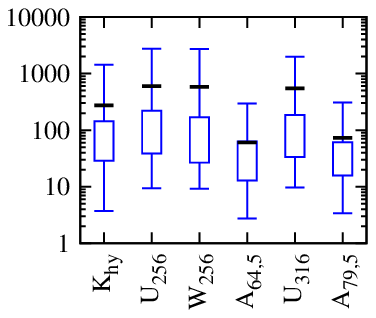}}%

\subfigure[landmark, $\epsilon=0.1$]{\label{fig:ag2_2_1} \includegraphics[width=0.24\textwidth]{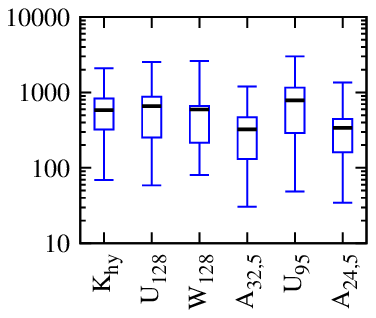}}%
\hfill
\subfigure[landmark, $\epsilon=1$]{\label{fig:ag2_2_2} \includegraphics[width=0.24\textwidth]{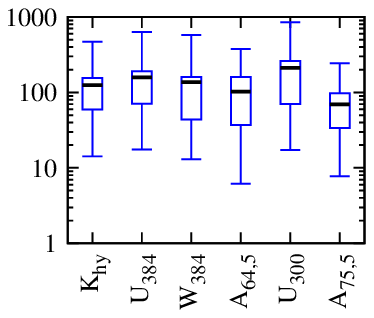}}%
\hfill
\subfigure[storage, $\epsilon=0.1$]{\label{fig:ag2_2_3} \includegraphics[width=0.24\textwidth]{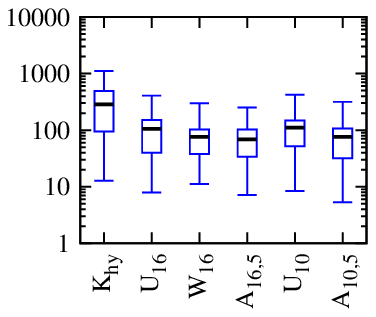}}%
\hfill
\subfigure[storage, $\epsilon=1$]{\label{fig:ag2_2_4} \includegraphics[width=0.24\textwidth]{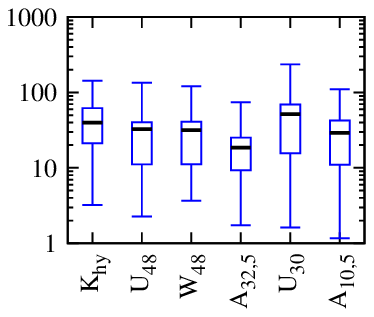}}%

	\caption{Comparing the absolute error of $6$ methods, from left to right, KD-hybrid, \UG with size giving lowest observed relative error, Privlet on this grid size, \AG with $m_1$ giving lowest observed relative error, \UG with suggested size, \AG with suggested size.  Here we use log scale because the absolute error have large ranges.} \label{fig:absolute}
\end{figure*}

In Figure~\ref{fig:compare} we perform an exhaustive comparison of $6$ methods: KD-hybrid, \UG with size giving lowest observed relative error, Privlet on this grid size, \AG with $m_1$ giving lowest observed relative error, \UG with suggested size, \AG with suggested size.  We use all $4$ datasets, and two $\epsilon$ values (0.1 and 1).  From these results, we observe that \AG consistently and significantly outperforms other methods.  We also observe that \UG with the suggested grid sizes provides about the same accuracy as KD-hybrid, and \AG with suggested grid sizes clearly outperforms all non-\AG methods.  When compared with \AG with the experimentally observed best grid size, the results are slightly worse but in general quite close.

In Figure~\ref{fig:absolute} we plot the same comparisons, but using absolute error, instead of relative error.  Here we use logscale for the candlesticks because the ranges of the absolute errors are quite large.  Again we observe that \AG methods consistently and significantly outperforms other methods.  It is interesting to note that for the road dataset, we observe that \UG  with suggested sizes outperform \UG using sizes optimized for the relative error.  Recall that this is the only dataset that has a large difference between suggested size and observed optimal size, because the dataset is highly uniform.  When one considers absolute errors, our suggest size seem to work very well.  This suggests the robustness of our error analysis and the guidelines that follow from the analysis.  Recall that our analysis did not depend upon the using of relative error or absolute error.

\comment{
\begin{figure*}
\subfigure[road, $\epsilon=0.1$]{\label{fig:ag1_2_1} \includegraphics[width=0.24\textwidth]{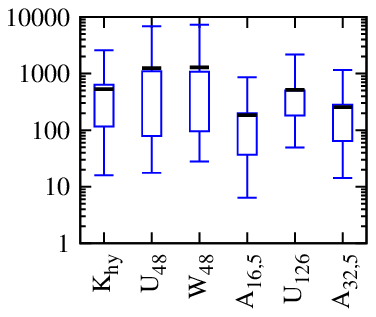}}%
\hfill
\subfigure[road, $\epsilon=1$]{\label{fig:ag1_2_2} \includegraphics[width=0.24\textwidth]{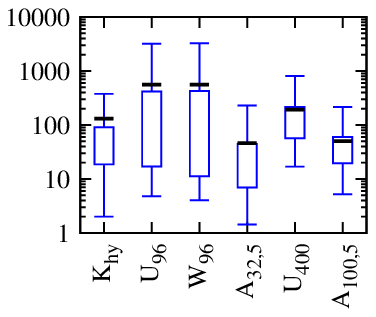}}%
\hfill
\subfigure[checkin, $\epsilon=0.1$]{\label{fig:ag1_2_3} \includegraphics[width=0.24\textwidth]{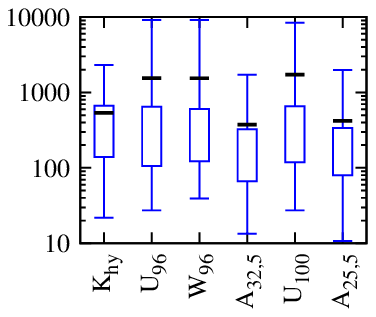}}%
\hfill
\subfigure[checkin, $\epsilon=0.1$]{\label{fig:ag1_2_4} \includegraphics[width=0.24\textwidth]{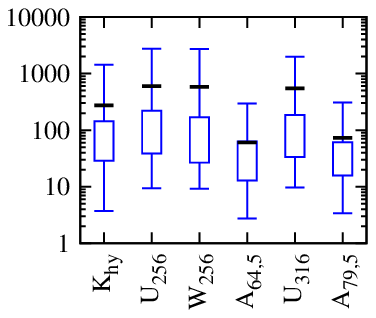}}%

\subfigure[landmark, $\epsilon=0.1$]{\label{fig:ag1_2_1} \includegraphics[width=0.24\textwidth]{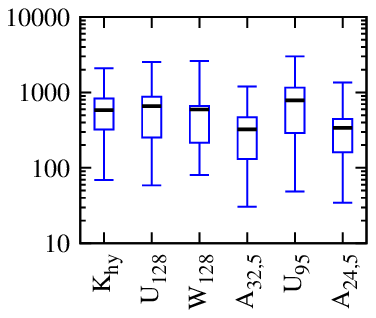}}%
\hfill
\subfigure[landmark, $\epsilon=1$]{\label{fig:ag1_2_2} \includegraphics[width=0.24\textwidth]{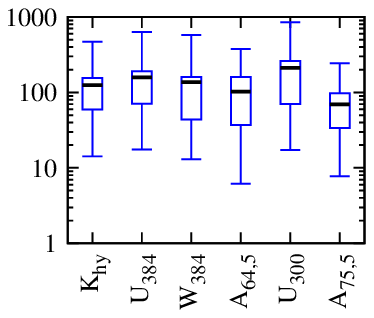}}%
\hfill
\subfigure[storage, $\epsilon=0.1$]{\label{fig:ag1_2_3} \includegraphics[width=0.24\textwidth]{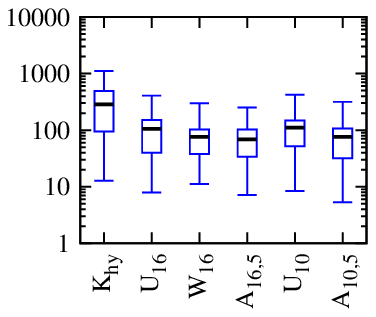}}%
\hfill
\subfigure[storage, $\epsilon=1$]{\label{fig:ag1_2_4} \includegraphics[width=0.24\textwidth]{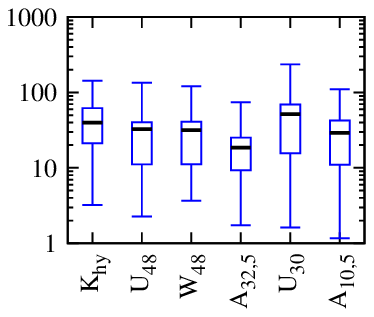}}%

	\caption{Comparing Absolute error of KD-hybrid, the best uniform partition method, wavelet method, the best hierarchical method and the best adaptive method} \label{fig:ag0}
\end{figure*}

\begin{figure*}
\subfigure[road, $\epsilon=0.1$]{\label{fig:ag1_2_1} \includegraphics[width=0.24\textwidth]{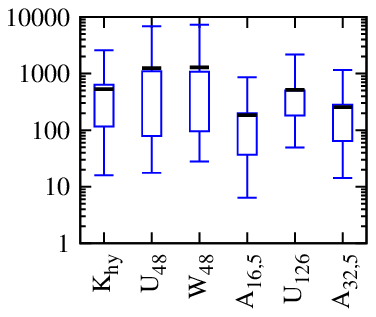}}%
\hfill
\subfigure[road, $\epsilon=1$]{\label{fig:ag1_2_2} \includegraphics[width=0.24\textwidth]{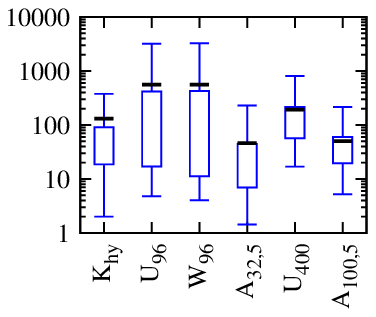}}%
\hfill
\subfigure[checkin, $\epsilon=0.1$]{\label{fig:ag1_2_3} \includegraphics[width=0.24\textwidth]{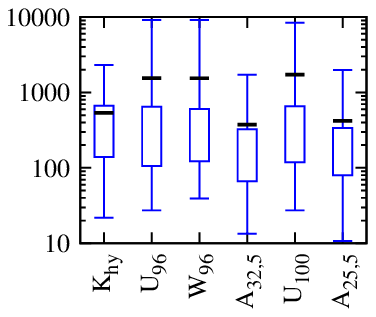}}%
\hfill
\subfigure[checkin, $\epsilon=0.1$]{\label{fig:ag1_2_4} \includegraphics[width=0.24\textwidth]{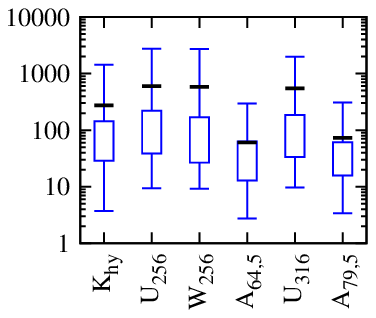}}%

\subfigure[landmark, $\epsilon=0.1$]{\label{fig:ag1_2_1} \includegraphics[width=0.24\textwidth]{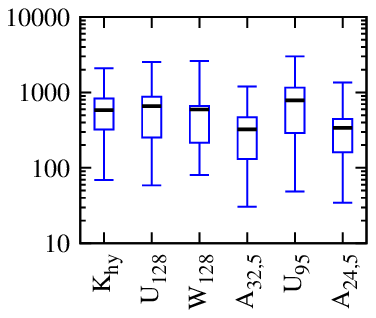}}%
\hfill
\subfigure[landmark, $\epsilon=1$]{\label{fig:ag1_2_2} \includegraphics[width=0.24\textwidth]{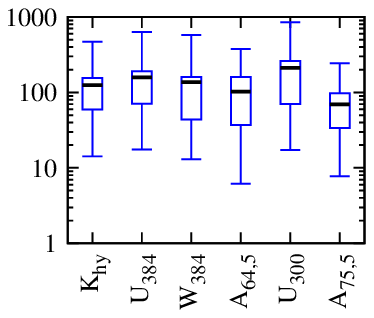}}%
\hfill
\subfigure[storage, $\epsilon=0.1$]{\label{fig:ag1_2_3} \includegraphics[width=0.24\textwidth]{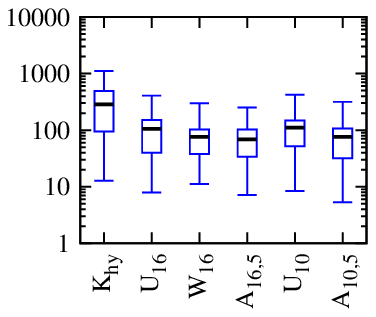}}%
\hfill
\subfigure[storage, $\epsilon=1$]{\label{fig:ag1_2_4} \includegraphics[width=0.24\textwidth]{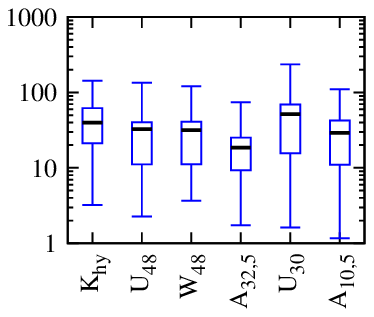}}%

	\caption{Comparing Absolute error of KD-hybrid, the best uniform partition method, wavelet method, the best hierarchical method and the best adaptive method} \label{fig:ag0}
\end{figure*}

\begin{table*}
\begin{center}
\begin{tabular}{|l|c|c|c|c|c|c|c|c|c|c|c|c|c|c|c|c|}
 \hline
 \multirow{3}{*}{dataset}
 & \multicolumn{8}{|c|}{$\epsilon=0.1$}
 & \multicolumn{8}{|c|}{ $\epsilon=1$}
 \\
 & \multicolumn{2}{|c|}{$K_{hy}$}
 & \multicolumn{2}{|c|}{$U$}
 & \multicolumn{2}{|c|}{$W$}
 & \multicolumn{2}{|c|}{$A$}

  & \multicolumn{2}{|c|}{$K_{hy}$}
 & \multicolumn{2}{|c|}{$U$}
 & \multicolumn{2}{|c|}{$W$}
 & \multicolumn{2}{|c|}{$A$}
 \\
 & {time} & {nodes} & {time} & {nodes}
 & {time} & {nodes} & {time} & {nodes}
 & {time} & {nodes} & {time} & {nodes}
 & {time} & {nodes} & {time} & {nodes}

 \\ \hline

 road & 79.28 & 320233 & 16.87 & 4608 & 15.06 & 4608 & 29.62 &17570
    & 63.29 & 270781 & 15.24&18432&16.03&18432&33.27&	170041
  \\ \hline
 checkin & 39.34 & 99265 & 10.23 & 18432 & 10.84 & 18432 &19.56 &12620
    & 42.41 & 141601 & 11.76&131072&19.76&131072&21.88&	111666

   \\ \hline
 landmark & 34.42 & 91459 & 9.33 & 32768 & 11.44 & 32768 & 16.81 &11917
    & 38.20 & 146191 & 12.45&294912&34.38&294912&	19.18&101425

      \\ \hline
 storage & 0.37 & 1339 & 0.08 & 512 & 0.12 & 512 & 0.17 &646
    & 0.47 & 2791 & 0.09&4608&0.34&4608&	0.19 &3048

\end{tabular}
\end{center}
\caption{Running time comparison}
\end{table*}
}

\comment{
Flat method is a fundamental method for 2-dimensional data. What the method do is equally partition the dataset into small cells and use whole privacy budget to answer queries. If a query does not include a whole cell, then use the ratio of noisy count to answer the query. We realize that flat method actually performs reasonably good and use flat method as our baseline.

We first compare flat method with the existing KD-tree method in \cite{sp12}. For the flat method, the only parameter is the partition size. We did experiment on both two dataset with different partition size and the result is presented in \ref{fig:kd1} and \ref{fig:kd2}.

From the figure we can learn that actually flat method performs pretty good. In most of the cases we tested, flat method can give a better result with a range of partition size.
} 
%\input{analysis}

%\section{Two Dimensional Publishing} \label{sec:?}
%\input{2dpub}

%\section{Discussions} \label{sec:discussions}
%\input{related}

\section{Conclusion} \label{sec:conclusions}

In this paper, we tackle the problem of releasing a differentially private synopsis for two dimensional datasets.  We have identified how to choose the partition granularity to balance errors due to two sources as the key challenge in differentially private synopsis methods, and propose a methodology for choosing grid size for the uniform grid method, based on an analysis of how the errors depend on the grid size.  
We have proposed a novel, simple, and effective adaptive grid method, together with methods for choosing the key parameters.  
We have conducted extensive evaluations using 4 real datasets, including large geo-spatial datasets that have not been used in differentially private data publishing literature before.  Experimental results validate our methodology and show that our methods outperform existing approaches.  
We have analyzed the effect of dimensionality on hierarchical methods, illustrating why hierarchical methods do not provide significant benefit in $2$-dimensional case, and predicting that they would perform even worse with higher dimensions.

%We first examine the current state of the art methods. Such methods essentially adapt binary partitioning, suitable for one dimensional datasets, to two dimensions. We argue that such methods are non-optimal. In addition, through extensive experimental comparisons on real datasets, we show that such methods generally perform no better than the basic uniform-grid approach. This approach applies an equi-width grid of a carefully chosen size over the data domain and then issues independent count queries on the grid cells.

%We further examine the uniform grid approach and show how to best choose the grid size. By analyzing how this method works, we show that its disadvantage is that it treats both dense and sparse regions in the same way. We thus introduce an adaptive grid method that exploits the need to have finer granularity partitioning over dense regions and, at the same time, coarse partitioning over sparse regions. The adaptive grid method lays a coarse-grained grid over the dataset, and then further partitions each cell according to its noisy count. Both levels of partitions are then used in answering queries over the dataset.

%To support the claims we make, we conduct extensive experimental evaluation using real-world datasets. We compare our approach to various methods including differentially private spatial indexes, binary hierarchical methods, as well as wavelet transform methods. Our results show that the adaptive grids methods performs consistently better than all the other approaches.

\bibliographystyle{IEEEtran}
\bibliography{privacy}

\end{document}